\documentclass[%
onecolumn,
 superscriptaddress,
nofootinbib,
 longbibliography,
 aps,
 prx
]{revtex4-2}
\usepackage{amsmath,amssymb,bm,braket,url,dsfont}
\usepackage{graphicx}
\usepackage{xcolor}
\usepackage{longtable}
\usepackage{multirow}
\usepackage{array}
\usepackage{booktabs}
\usepackage{ulem}
\newcolumntype{L}{>{$}l<{$}} % left-aligned column with math mode
\newcolumntype{C}{>{$}c<{$}} % left-aligned column with math mode

\usepackage[
draft=false,
bookmarks=true,
bookmarksnumbered=false,
bookmarksopen=false,
bookmarksopenlevel=4,
colorlinks=true,
anchorcolor=black,
citecolor=blue,
filecolor=magenta,
linkcolor=blue,
linkbordercolor={1 0 0},
urlcolor=blue,
pdfborder={0 0 1},
pdftitle={},
pdfauthor={},
pdfsubject={},
pdfkeywords={},
pdfpagemode=UseThumbs,
]{hyperref}

\newcommand{\cota}{{CoTa$_3$S$_6$}}
\newcommand{\sop}[2]{\left( #1,#2\right)}
\newcommand{\sping}{\mathcal{G}}
\newcommand{\msg}{\mathbf{G}}
\newcommand{\spog}{\mathcal{P}_\text{sp}}
\newcommand{\spinpg}{\mathcal{P}}
\newcommand{\mpg}{\mathcal{P}_\text{orb}}
\newcommand{\mpgsoc}{\mathbf{P}}
\newcommand{\spgop}[2]{^{#1}#2}
\newcommand{\spdim}{\mathcal{D}_\text{sp}}

\newcommand{\mq}{\mathrm{Q}}
\usepackage{orcidlink}

\graphicspath{%
{./images}%
}

\begin{document}

\title{
Symmetry analysis with spin crystallographic groups:\\
Disentangling effects free of spin-orbit coupling in emergent electromagnetism
}

\author{Hikaru Watanabe  \orcidlink{0000-0001-7329-9638}} % https://orcid.org/0000-0001-7329-9638
\email{hikaru-watanabe@g.ecc.u-tokyo.ac.jp}
\affiliation{Research Center for Advanced Science and Technology, University of Tokyo, {Meguro-ku}, Tokyo 153-8904, Japan}

\author{Kohei Shinohara  \orcidlink{0000-0002-5907-2549}} % https://orcid.org/0000-0002-5907-2549
\affiliation{Department of Materials Science and Engineering, Kyoto University, {Sakyo}, Kyoto 606-8501, Japan}

\author{Takuya Nomoto  \orcidlink{0000-0002-4333-6773}} % https://orcid.org/0000-0002-4333-6773
\affiliation{Research Center for Advanced Science and Technology, University of Tokyo, {Meguro-ku}, Tokyo 153-8904, Japan}
% \affiliation{CREST, Japan Science and Technology Agency, {Saitama} 332-0012, Japan}

\author{Atsushi Togo  \orcidlink{0000-0001-8393-9766}} % https://orcid.org/0000-0001-8393-9766
\affiliation{Center for Basic Research on Materials, National Institute for Materials Science, {Tsukuba}, Ibaraki 305-0047, Japan}

\author{Ryotaro Arita  \orcidlink{0000-0001-5725-072X}} % https://orcid.org/0000-0001-5725-072X
\affiliation{Research Center for Advanced Science and Technology, University of Tokyo, {Meguro-ku}, Tokyo 153-8904, Japan}
\affiliation{RIKEN, Center for Emergent Matter Science, {Saitama} 351-0198, Japan}

\begin{abstract}
Recent studies identified spin-order-driven phenomena such as spin-charge interconversion without relying on the relativistic spin-orbit interaction.
Those physical properties can be prominent in systems containing light magnetic atoms due to sizable exchange splitting and may pave the way for realization of giant responses correlated with the spin degree of freedom.
In this paper, we present a systematic symmetry analysis based on the spin crystallographic groups and identify the physical property of a vast number of magnetic materials up to 1500 in total.
By decoupling the spin and orbital degrees of freedom, our analysis enables us to take a closer look into the relation between the dimensionality of spin structures and the resultant physical properties and to identify the spin and orbital contributions separately.
In stark contrast to the established analysis with magnetic space groups, the spin crystallographic group manifests richer symmetry including spin translation symmetry and leads to emergent responses.
For representative examples, we discuss geometrical nature of the anomalous Hall effect and magnetoelectric effect and classify the spin Hall effect arising from the nonrelativistic spin-charge coupling.
Using the power of computational analysis, we apply our symmetry analysis to a wide range of magnets, encompassing complex magnets such as those with noncoplanar spin structures as well as collinear and coplanar magnets.
We identify emergent multipoles relevant to physical responses and argue that our method provides a systematic tool for exploring sizable electromagnetic responses driven by spin order.
\end{abstract}

\maketitle

\section{introduction}

Spintronics has experienced tremendous growth, and the concept has been discussed in various fields including topological electronic systems and superconductors.
In recent years, spin-orbit coupling (SOC), a relativistic interaction between the charge and spin degrees of freedom, is particularly of matter due to its rich physical consequences.
Search for candidate materials has taken place to maximize the physical responses associated with spin-orbit interaction in these decades.
For example, giant spin-momentum splittings have been identified in systems with heavy atoms having large SOC.
Their strong spin-orbit entanglement has been demonstrated by spectroscopy~\cite{Ishizaka2011-dz,Krempasky2016-to} and transport measurements~\cite{Ideue2021-wi}.
The progress may let us consider the possibility of physics originating from SOC covering a broader range of materials other than what consists of heavy atoms, such as materials based on $3d$ transition metal elements with negligible relativistic SOC.  

To this end, a concept of nonrelativistic spin-charge locking has been proposed in theories~\cite{Hayami2019-vh,Yuan2020-kx,Yuan2021-el,Smejkal2022-ir}.
This coupling arises from the spontaneous magnetic ordering without the help of SOC and thereby can exhibit exchange splitting energy comparable to that of the Coulomb interaction.
The concept illuminates the potential impacts of light elements for the spintronic application and further identified advantageous properties compared to conventionally-studied materials; \textit{e.g.}, strong exchange splitting energy and large transition temperature.
Notably, the magnetic order gives rise to characteristic spin-momentum-locking structure due to coupling between the order and structural property of crystals as found in antiferromagnetic materials.
These aspects are valuable for applications in the field of antiferromagnetic spintronics gathering considerable interest as an emerging field in condensed matter physics~\cite{Baltz2018-pu,Manchon2019-xj,Smejkal2022-ao}; for instance, various physical phenomena free from SOC have been clarified in the previous works such as the spin-polarized current induction~\cite{Zelezny2017-tf,Zhang2018-yj,Naka2019-pf,Naka2021-ow,Gonzalez-Hernandez2021-pb}, nonlinear response~\cite{Hayami2022-lb,Hayami2022-wt}, piezomagnetic effect~\cite{Hayami2019-vh,Ma2021-ji,Bhowal2022-ui}, and magnetoresistance~\cite{Smejkal2022-zw}.

Released from the SOC constraint, the array of localized spins does not have any favorable orientation described by the crystal structure.
The decoupling between spin and orbital degrees of freedom leads to a magnetic symmetry higher than the conventional magnetic space group symmetry (Shubnikov group).
Such magnetic symmetry without SOC is covered by the \textit{spin crystallographic group} such as spin space group and spin point group~\cite{Brinkman1966-qg,Opechowski1986} which includes a richer group structure due to the absence of SOC.
The spin crystallographic groups have applied to analyzing the electronic structure modified by the spontaneous spin order, particularly in the case of simple spin configurations~\cite{Hayami2019-vh,Ahn2019-cd,Yuan2020-kx,Liu2022-dh,Smejkal2022-ir,Smejkal2022-ga}.
For instance, recent studies identified a series of spin crystallographic symmetries providing the nonrelativistic spin-charge coupling by which the degeneracy at each crystal momentum is lifted.
The materials manifesting such spin crystallographic symmetry are characterized by a collinear antiferromagnetic structure whose magnetic unit cell is the same as the chemical one due to the zero propagation vector.
Candidate materials such as MnTe for the so-called altermagnets belong to this class~\cite{Smejkal2022-ao,Smejkal2022-ga,Krempasky2024-ve,Reimers2024-kh,Osumi2024-tb,Lee2024-hf}.

One can expect that there exist rich physical consequences of nonrelativistic spin-charge coupling in other kinds of collinear magnets as well as more complex spin-ordered systems (\textit{e.g.,} spin structure with nonzero propagation vectors), which have not been rarely investigated from the viewpoint of spin crystallographic group.
The latter class encompasses intriguing systems such as noncollinear,  noncoplanar, and multiferroic magnets~\cite{Zelezny2017-tf}.
A prototypical phenomenon unique to these materials is the geometrical Hall effect~\cite{Nagaosa2010-dn}.
The effect occurs in magnets with noncoplanar spin structure observed in materials such as systems with a triangular or  net.
The resultant time-reversal-symmetry breaking resembles the orbital-flux order proposed in Ref.~\cite{Haldane1988-dm}, and does not require the relativistic SOC effect~\cite{Ohgushi2000-tf}.
The SOC-free nature is in high contrast to the well-known anomalous Hall effect arising from the collinear and coplanar spin order~\cite{Ye1999-ik,Ohgushi2000-tf,Shindou2001} and may be responsible for the anomalous Hall responses of magnetic skyrmion crystals~\cite{Neubauer2009-ya,Fujishiro2021-ri}.
These prior studies indicate that the dimension of the spin structure is key to identifying the emergent physical responses induced by the spin order without the help of SOC.
In this regard, the spin crystallographic group is advantageous compared to the widely-used magnetic space group, because its group symmetry reflects a given spin-structure dimension.

In this paper, we present the spin-crystallographic-group symmetry analysis covering not only the simple magnetic materials having collinear and coplanar spin structures with the zero propagation vector but also complex spin-ordered systems such as noncoplanar magnets and those with non-zero propagation vectors.
The analysis incorporates the dimensionality of the spin structure and hence provides a convenient tool for identifying the emergent symmetry breaking and associated phenomena which cannot be distinguished from the SOC-assisted contribution in terms of the SOC-accounted symmetry analysis based on magnetic space group and magnetic point group~\cite{nye1985physical}.

Specifically, we demonstrate the following points of the present symmetry analysis; by separating the spin and orbital spaces, we can identify the contribution of each degree of freedom to physical responses.
Our symmetry analysis identified nontrivial spin crystallographic symmetry where the spin-space symmetry is kept highly symmetric to be unexpected by its crystal structure; \textit{e.g.,} the cubic spin-space symmetry despite the axial symmetry of the crystal.
Despite the inactive spin-related quantities, the symmetry does not forbid physical phenomena involving the orbital degree of freedom such as the geometrical Hall effect.
Such nontrivial spin crystallographic symmetry enables us to explore the spin-geometry-induced response so as to unambiguously distinguish it from the relativistic SOC effect.
The orbital-active but spin-inactive aspects highlight the significance of magnets with non-zero propagation vectors and are in sharp contrast to the previously identified SOC-free physical property, that is spin-active but orbital-inactive property~\cite{Zhang2018-yj}.
Furthermore, our result suggests that emergent physical responses can be examined in a semi-quantitative manner by combining the spin-crystallographic-group symmetry analysis with the physical insights into spin fluctuations correlated with the dimensionality of a spin structure.
We explain these features by taking several physical properties such as the anomalous Hall effect, magnetoelectric effect, and spin Hall effect.
The symmetry analysis is computationally performed with the use of the algorithm for searching the spin space group developed by Shinohara \textit{et al.}~\cite{Shinohara2024-ld}. 
By classifying a vast number of magnetic materials ($\sim$ 1500), we systematically clarify the SOC-free emergent properties of real magnetic materials.

\begin{figure*}[htbp]
                \centering
                \includegraphics[width=0.75\linewidth,clip]{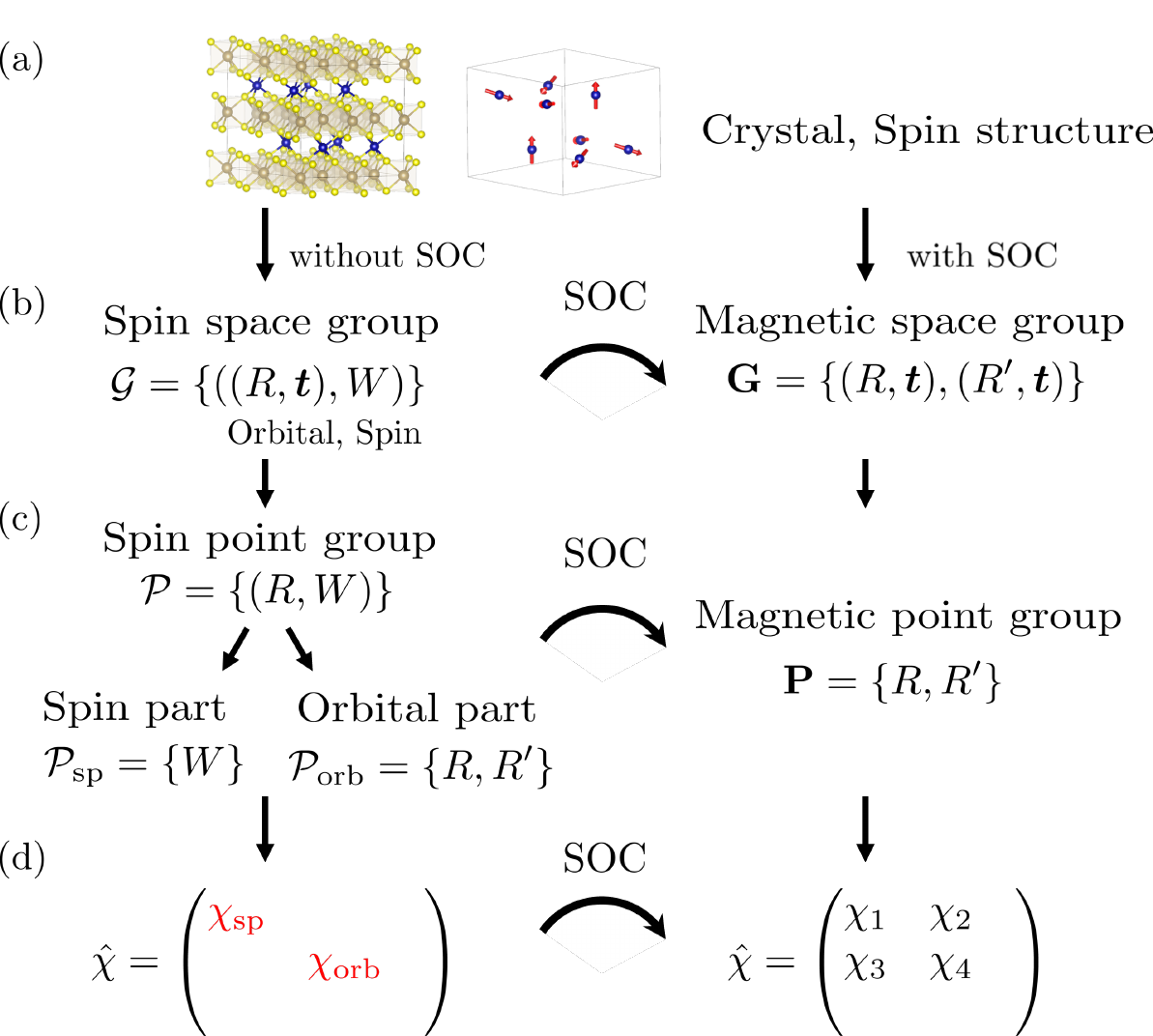}
                \caption{
                Procedure of symmetry analysis.
                (a) Input of crystal and spin structures. Dozens of data imported from \textsc{magndata}.
                (b) Spin space group $\sping$ and magnetic space group $\msg$ computationally identified. The effect of spin-orbit coupling (SOC) is not considered in $\sping$ but is taken into account in $\msg$.
                The spin-space-group symmetry is given by the spin rotation ($W$) and by orbital-space operations such as rotation ($R$) and translation ($\bm{t}$) operations.
                The magnetic space group is comprised of the orbital-space operations $(R,\bm{t})$ with or without the time-reversal operation $\theta$ such as $R' = \theta R$.
                For the magnetic space group $\msg$, the orbital-space operations $R,R'$ also act on spins due to the SOC constraint.
                (c) The space group is reduced to its point group by omitting the translation operation $\bm{t}$.
                Spin ($\spinpg$) and magnetic ($\mpgsoc$) point groups are obtained from $\sping$ and $\msg$, respectively.
                The spin point group is further divided into ($\spog, \mpg$) consisting of either spin- or orbital-space symmetry.
                The spin crystallographic groups ($\sping,\spinpg$) are reduced to the spin-orbital-coupled groups ($\msg,\mpgsoc$) by SOC.
                (d) The symmetry-adapted form of a given tensor $\hat{\chi}$ obtained by the spin or magnetic point group symmetry.
                The red-colored components ($\chi_\text{sp},\chi_\text{orb}$) originates from the spin-order-induced symmetry breaking without SOC, and the origin of each component is further attributed to the spin ($\chi_\text{sp}$) and orbital ($\chi_\text{orb}$) degree of freedom.
                SOC entangles the spin contribution with the orbital ($\chi_{1},\chi_{4}$) and induces additional components ($\chi_{2},\chi_{3}$).
                }
                \label{Fig_schematics}
                \end{figure*}

The organization of the paper is the following.
In Sec.~\ref{Sec_spin_group_basic}, we overview the spin space group and introduce the symmetry analysis with it.
Based on the spin point group, the symmetry analysis is applied to some of ferromagnetic and antiferromagnetic materials and their physical responses in Sec.~\ref{Sec_classify_physical_properties}.
Section~\ref{Sec_magndata_study} is devoted to a high-throughput symmetry analysis of magnetic materials listed in \textsc{magndata}~\cite{Gallego2016-yz,Gallego2016-pb}.
In light of emergent magnetic multipoles and rotators for spin-charge coupling, we investigate the SOC-free physical properties and demonstrate that our symmetry analysis clarifies the importance of the spin-structure dimension.
We summarize the contents in Sec.~\ref{Sec_discussions}.
The procedure of the symmetry analysis is sketched in Fig.~\ref{Fig_schematics}.

The symmetry analysis with given spin space group $\sping$ and magnetic space group $\msg$ are automatically computed on the basis of \textsc{spglib}~\cite{togo2018tspglib,Shinohara2023-sy} and \textsc{spinspg}~\cite{Shinohara2024-ld}.
Terms on group theory can be found in Appendix~\ref{SecApp_note_group_theory}.

\section{Spin-group operations and spin crystallographic group}
\label{Sec_spin_group_basic}

We introduce the transformation property of symmetry operations and overview the spin crystallographic symmetry such as the spin space group and spin point group.
We not only explain the spin space group but also introduce the spin and orbital parts of the spin crystallographic group to disentangle the spin and orbital contributions to physical phenomena.
We concisely introduce the notations of spin crystallographic groups we adopt, while the terminology related to the group theory and its mathematical aspect are summarized in Appendix~\ref{SecApp_note_group_theory}.
Concerning the crystallographic property, interested readers can refer to Refs.~\cite{Opechowski1986,Litvin1974-gr}.

Owing to the absence of generic spin-orbital coupling, the rotation operation separately acts on the spin and orbital space in terms of the spin group symmetry.
Let $g= \sop{R}{W}$ be the combination of symmetry operations $R$ and $W$ acting on the orbital and spin space, respectively.

\begin{figure*}[htbp]
                \centering
                \includegraphics[width=0.90\linewidth,clip]{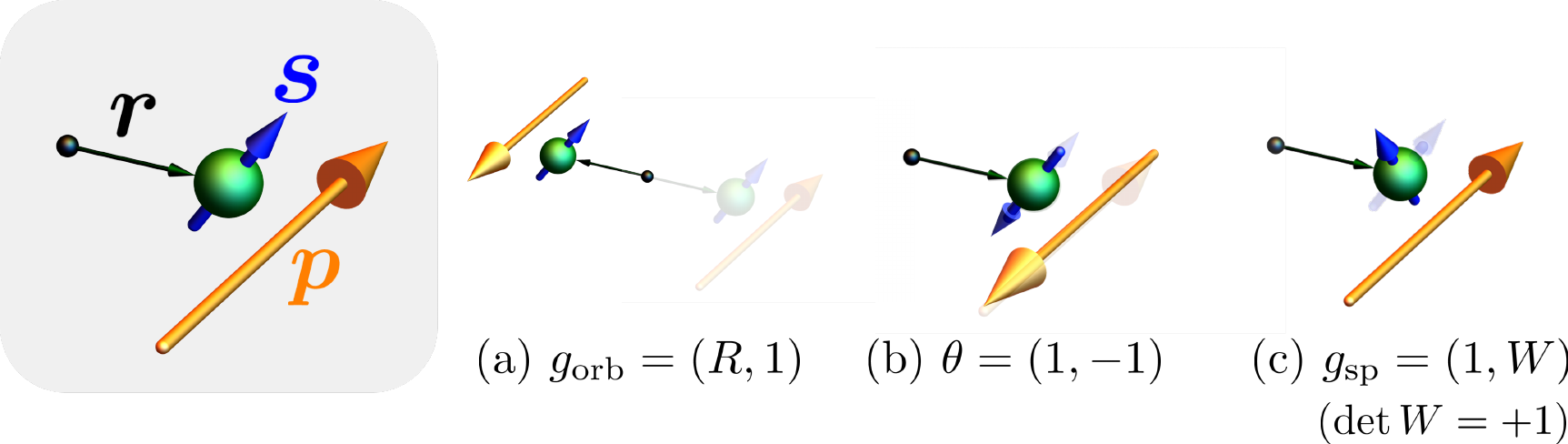}
                \caption{
                Spin-group transformation of electron depicted by position $\bm{r}$, momentum $\bm{p}$, and spin $\bm{s}$.
                The spin-group operation $g = \sop{R}{W}$ ($R$ is orbital-space rotation, $W$ is spin-space rotation).
                (a) Orbital-space operation acting on position and momentum while leading to no action on spins ($W = 1$).
                (b) Spin-space inversion operation same as the time-reversal operation ($R = 1,~W = -1$) flipping the time-reversal-odd quantities such as $\bm{r}$ and $\bm{s}$.
                (c) Spin-space operation satisfying $R = 1$ and the proper rotation condition (det\,$W=+1$).
                It gives no transformation related to the orbital-space objects.
                }
                \label{Fig_symops}
                \end{figure*}

Firstly, we consider the point group symmetry with rotation operations $\sop{R}{W}$ and raise some examples of the basic transformation property such as position $\bm{r}$, momentum $\bm{p}$, and spin $\bm{s}$.
For instance, let us take the orbital-space-only operation $g_\text{orb} = \sop{R}{1}$, the time-reversal operation $\theta = \sop{1}{-1}$, and spin-space proper rotation $g_\text{sp} = \sop{1}{W}$ (det\,$W= + 1$).
Note that the time-reversal operation is denoted by the space-inversion operation in the spin space by following Refs.~\cite{Litvin1977,Shinohara2024-ld}.
The operators are transformed under each symmetry operation as shown in Fig.~\ref{Fig_symops}.
When the improper property holds for the spin-space operation as det\,$W= - 1$, the operation $g$ includes the time-reversal operation $\theta$ such as that with the spin-space space-inversion ($W=-1$) and mirror operation ($W = m$).
It follows that $g$ with det\,$W= + 1$ (det\,$W= - 1$) is unitary (anti-unitary). 
Owing to the time-reversal operation $\theta$, the operation $g$ with det\,$W= - 1$ can act on the orbital space as it flips the time-reversal-odd quantities, \textit{e.g.}, $\bm{p}$ [Fig.~\ref{Fig_symops}(b)] and orbital magnetization.
On the other hand, the proper rotation in the spin space does affect only the spin degree of freedom [Fig.~\ref{Fig_symops}(c)].
For a spin-group operation $g = \sop{R}{W}$, the operators are transformed as
                \begin{align}
                g \, r_a \,g^{-1} &=  r_b T_{ba} (R) ,\\
                g \, p_a \,g^{-1}  &=   \text{det} W \cdot  p_b T_{ba} (R) ,\\
                g \, s_a \,g^{-1}  &=  s_b T_{ba} (W) ,
                \end{align}
where we introduced the three-dimensional orthogonal matrices $\hat{T} (u)$ ($u \in \text{O(3)}$) in accordance with the vectorial symmetry of each object such as $\hat{T} (-1) = - \mathbf{1}$ for the inversion operation ($\mathbf{1}$ is the identity matrix in three-dimensional system)~\footnote{
        Note that the representation matrices are in the Cartesian coordinates though they are usually in the basis spanned by the Bravais vectors in the field of crystallography.
        This is because the spin-space operations do not necessarily belong to the Bravais class same as that for a given crystal structure in the framework of spin crystallographic group.
}.

We generalize the symmetry argument to the case of the tensor quantity $O_{abc\cdots}$.
The transformation is written by
        \begin{align}
            g\, O_{abc\cdots} \, g^{-1}
            &=  O_{a'b'c'\cdots} D_{a'a}^{(A)} (g)D_{b'b}^{(B)} (g) D_{c'c}^{(C)} (g)  \cdots,
            \label{tensor_transformation}
        \end{align}
where we introduced the representation matrices for physical quantities $A,B,C,\cdots$ labeled by the indices $a,b,c,\cdots$, respectively.
For the aforementioned three quantities, the representation matrices are explicitly given by
                \begin{equation}
                \hat{D}^{(r)} (g) = \hat{T} (R),~\hat{D}^{(p)} (g) = \text{det}\,W \cdot \hat{T} (R),~\hat{D}^{(s)} (g) = \hat{T} (W).
                \label{representation_matrices_examples}
                \end{equation}
Taking the operations depicted in Fig.~\ref{Fig_symops}, the representation matrices are explicitly given by
                \begin{align}
                &\hat{D}^{(r)} (g_\text{orb}) = \hat{T} (R),~\hat{D}^{(p)} (g_\text{orb}) =   \hat{T} (R),~\hat{D}^{(s)} (g_\text{orb}) = \mathbf{1},\\
                &\hat{D}^{(r)} (\theta) = \mathbf{1},~\hat{D}^{(p)} (\theta) =   - \mathbf{1},~\hat{D}^{(s)} (\theta) = \hat{T} (W),\\
                &\hat{D}^{(r)} (g_\text{sp}) = \mathbf{1},~\hat{D}^{(p)} (g_\text{sp}) =  \mathbf{1},~\hat{D}^{(s)} (g_\text{sp}) = \hat{T} (W),
                \end{align}
by which, for instance, the spin-space mirror operation ($R=1, W=m$) is obtained by combining the representation matrices of $\theta$ with that of $g_\text{sp}$.
Then, in the absence of SOC, the transformation property of spins under $W$ can be given similarly to that of a polar vector.
Owing to the irrelevant role of proper spin-space rotations in the transformation of $\bm{r}$ and $\bm{p}$ [Eq.~\eqref{representation_matrices_examples}], the system has the \textit{orbital time-reversal symmetry} if there exists a symmetry operation given by $g = \sop{1}{W}$ (det\,$W=-1$).

Next, we consider the structure of the spin crystallographic group.
In terms of the space group symmetry, the orbital-space operation $h$ is comprised of the point group operation $R$ and the translation operation $\bm{t}$ as $h = \left( R,\bm{t} \right)$.
The \textit{spin space group} $\sping$ is a set of symmetry operations $g = \sop{h}{W}$ under which crystal structures and spin configuration dwelling on each magnetic atom are invariant.
In stark contrast to the well-known magnetic space group (Shubnikov group) $\msg$~\cite{litvin2014magnetic}, we can take symmetry operations acting on objects in the orbital and spin space independently~\cite{Litvin1974-gr}.
The difference can be inferred from the adopted Hamiltonian as follows.

Let us consider Hamiltonian manifesting the spin-space-group symmetry.
The Hamiltonian not only consists of kinetic and potential Hamiltonians $H_0$ for paramagnetic states but also takes into account the spontaneous spin ordering by the molecular field.
The total Hamiltonian is given by
        \begin{equation}
        H_\text{SG} = H_0 + H_\text{mag},
        \label{hamiltonian_for_spin_space_group}
        \end{equation}
where the molecular-field term is 
        \begin{equation}
            H_\text{mag} = \sum_i \bm{B}^{(i)} \cdot \bm{s}^{(i)},
            \label{spin_order_molecular_field}
        \end{equation}
with indices for the sites $i$.
The exchange-splitting field $\bm{B}^{(i)}$ is defined at each magnetic site ($\bm{B}^{(i)} = \bm{0}$ for nonmagnetic atoms).
The Hamiltonian [Eq.~\eqref{hamiltonian_for_spin_space_group}] is invariant under the associated spin space group as
        \begin{equation}
        g \in \sping,~g\, H_\text{SG}\, g^{-1} = H_\text{SG}.
        \label{spin_space_group_relation}
        \end{equation}
The paramagnetic part satisfies the following relation for $g = \sop{h}{W} \in \sping$ as
        \begin{equation}
                H_0 = \sop{h}{W} H_0 \sop{h}{W}^{-1} = \sop{h}{1} H_0 \sop{h}{1}^{-1},
        \end{equation}
where the spin rotation $W$ is irrelevant.
The orbital-space part $h$ is therefore restricted by the atomic configuration and generated by the space group of a given crystal structure.
On the other hand, the exchange Hamiltonian is transformed under both $h$ and $W$ as
        \begin{align}
            g \left( \sum_{i} B_a^{(i)} s_a^{(i)}  \right) g^{-1} 
                &= \sum_i B_a^{(i)} \cdot \left(  g\, s_a^{(i)} \, g^{-1} \right) , \\
                &= \sum_{i}   B_a^{(i)} \cdot  s_b^{(i_h)} T_{ba} (W).
                \label{soc_transformation_under_sping}
        \end{align}
The orbital-space operation permutes the sites as $h : \bm{r}_i \mapsto \bm{r}_{i_h} = h^{-1} \,\bm{r}_i \,h$.
Owing to Eq.~\eqref{spin_space_group_relation}, the spin-group operations should satisfy
                \begin{equation}
                \hat{T} (W) \, \bm{B}^{(i)} =  \bm{B}^{(i_h)},
                \label{spin_space_group_restriction_magfield}
                \end{equation}
for every site, and the coupling arises between the orbital and spin degrees of freedom.
As the result, the spin-space operations are determined by $H_\text{mag}$, whereas orbital-space operations are by the total Hamiltonian.
Importantly, the operations $h$ and $W$ are taken independently as long as they satisfy Eq.~\eqref{spin_space_group_restriction_magfield}.
The property clearly distinguishes the spin space group from the magnetic space group.
We note that the spin-space-group symmetry holds in general if one properly takes into account the spin order in a SOC-free manner, while the spin-ordering effect is simply taken as the molecular field for illustrative purposes.

For the case of the magnetic space group, we similarly treat the magnetic order as the molecular fields and add spin-orbit interaction $H_\text{SOC}$ to the Hamiltonian.
The Hamiltonian reads as
        \begin{equation}
        H_\text{MG} = H_0 + H_\text{mag} + H_\text{SOC},
        \label{hamiltonian_for_magnetic_space_group}
        \end{equation}
where the SOC Hamiltonian is given by
        \begin{align}
        H_\text{SOC} &= \lambda \bm{L} \cdot \bm{s}.
        \label{SOChamiltonian}
        \end{align}
$\bm{L}$ denotes the atomic orbital angular momentum and $\lambda$ is the strength of SOC.
The additional symmetry constraint by the SOC Hamiltonian leads to the group-subgroup relation $\sping$ ($\msg < \sping$).
In sharp contrast to the spin space group $\sping$, the SOC Hamiltonian imposes the following constraint on $\sop{h}{W} = \sop{\left( R,\bm{t} \right)}{W} \in \msg$,
        \begin{equation}
         \text{det} R \cdot R = \text{det} W \cdot W.
         \label{SOC_constraint}
        \end{equation}
That is, the proper rotation parts of $h$ and $W$ should be the same as each other~\cite{Shinohara2024-ld}.
Eq.~\eqref{SOC_constraint}, along with Eq.~\eqref{spin_space_group_restriction_magfield}, ties the orbital space with the spin space.

Then, let us overview the group structure of spin space group $\sping$ which has been investigated in Refs.~\cite{Litvin1973-ii,Litvin1974-gr}.
$\sping$ contains the spin-only group $\sping_\text{so}$ as a normal subgroup ($\sping \triangleright \sping_\text{so} $).
The spin-only group solely consists of the spin symmetry operations such as 
        \begin{equation}
                \sping_\text{so} = \sop{(1,\bm{0})}{\spinpg_\text{so} }  = \{ \sop{(1,\bm{0})}{W} |~ W \in \spinpg_\text{so}\}.
        \end{equation}
The group is determined by the dimension of a given spin configuration, which we call the spin-structure dimension $\spdim$~\cite{Opechowski1986}.
For one-dimensional magnets (collinear magnets) denoted by $\spdim = 1$, the magnetic moments are parallel or anti-parallel to the axis $\bm{n}$ in the spin space.
The spin-only group is given by an internal semidirect product,
        \begin{equation}
            \spinpg_\text{so}  = \text{SO(2)} \rtimes \{1, m_\parallel \},
            \label{spin_only_group_1dim}
        \end{equation}
by which the vector $\bm{n}$ is invariant.
We can take rotation operations along $\bm{n}$ with an arbitrary angle and mirror reflection $m_\parallel$ whose mirror plane contains the axis $\bm{n}$. 
For the two-dimensional case ($\spdim = 2$, noncollinear but coplanar), the spin-only group is
        \begin{equation}
            \spinpg_\text{so}  = \{1, m_\perp \}.
            \label{spin_only_group_2dim}
        \end{equation}
The mirror operation $m_\perp$ shares its plane with the spins spanning the two-dimensional plane.
Lastly, in the three-dimensional case ($\spdim = 3$, noncoplanar), the spin-only group trivially consists of only the identity operation;
        \begin{equation}
            \spinpg_\text{so}  = \{1 \}.
            \label{spin_only_group_3dim}
        \end{equation}
Note that the orbital time-reversal symmetry is preserved in collinear magnets as well as coplanar magnets due to the spin-space mirror operation.
By using the spin-only group, the spin space group is decomposed as~\cite{Litvin1973-ii}
        \begin{equation}
        \sping  = \sping_\text{so}  \times \overline{\sping}.
        \label{spingroup_structure}
        \end{equation} 
As a result, we obtain the nontrivial spin space group $\overline{\sping}$ whose spin-space operation $W$ is intimately coupled to the orbital-space operation $h = (R,\bm{t})$ such as the combined operation of spin rotation and translation $g = \sop{(1,\bm{t})}{W}$;\textit{ i.e.,} the symmetry operation $\sop{(R,\bm{t})}{W}$ with $W \neq 1$ satisfies $(R,\bm{t}) \neq (1,\bm{0})$.

The set of spin translation operations $\{\sop{(1,\bm{t})}{W}\}$ in $\overline{\sping}$ forms the group which we denote the nontrivial spin translation group $\overline{\sping}_\text{st}$.
When the spin order does not modify the paramagnetic unit cell, the nontrivial spin translation group is reduced to the translation group $\mathcal{T} = \{ \sop{(1,\bm{t})}{1}\}$.
The nontrivial spin translation group is a normal subgroup of the nontrivial spin space group ($\overline{\sping} \, \triangleright \, \overline{\sping}_\text{st}$).
Thus, we decompose $\overline{\sping}$ by $\overline{\sping}_\text{st}$ as
                \begin{equation}
                \overline{\sping} = \bigcup_{i} ~g_i \, \overline{\sping}_\text{st}.
                \end{equation}
The representative $g_i = \sop{\left( R,\bm{t} \right)}{W}$ indicates that the spin-space operation is coupled to the orbital-space point group operations otherwise it is the identity ($W=1$); in other words, the orbital-space point group operation in $g_i$ is nontrivial ($R\neq 1$) except for the identity operation $\sop{(1,\bm{0})}{1}$.
Note that one can obtain another decomposition of the spin space group by the spin translation group defined by $\sping_\text{st} = \sping_\text{so} \times \overline{\sping}_\text{st}$ (see also Appendix~\ref{SecApp_note_group_theory}).
To corroborate the effect of the spin-structure dimension $\spdim$ on emergent responses, we here utilize the decomposition of Eq.~\eqref{spingroup_structure} in the following.
The spin space group $\sping$ for given crystal and spin structures can be computationally obtained~\cite{Shinohara2024-ld}.

For instance, we consider the body-centered-cubic Fe (space group $Im\bar{3}m$, No.~229) whose ferromagnetic spin polarization is along the $[001]$ axis.
The magnetic space group $\msg = I4/mm'm'$ indicates the spontaneous crystalline symmetry reduction from the cubic to tetragonal under the SOC effect.
On the other hand, the spin space group $\sping$ retains high symmetry in the ordered state.
The spin space group comprises the spin-only group $\sping_\text{so}$ given by Eq.~\eqref{spin_only_group_1dim} with the spin-space axis $\bm{n} = [001]$.
By dividing $\sping$ by the spin-only group, we obtain the nontrivial spin space group $\overline{\sping}$ [Eq.~\eqref{spingroup_structure}].
Since the ferromagnetic order does not modify the unit cell, the spin translation group is the translation group $\mathcal{T}_\text{bcc}$ for the bcc centering.
Then, the nontrivial spin space group is
                \begin{equation}
                \overline{\sping} = \bigcup_{i} ~g_i \, \overline{\sping}_\text{st} = \bigcup_{i} ~g_i \mathcal{T}_\text{bcc},
                \end{equation}
where $\{g_i\}$ forms the cubic point group same as that for the paramagnetic state ($m\bar{3}m$).
As a result, the nontrivial spin space group is the same as paramagnetic one $\overline{\sping} =Im\bar{3}m$ (each spin-space operation is the identify operation and hence omitted), and the overall spin space group symmetry is given by
                \begin{equation}
                \sping  = \left\{ \sop{h}{W} |~ h \in Im\bar{3}m,~W \in  \text{SO(2)} \rtimes \{1, m_\parallel \} \right\}. 
                \label{Fe_FM_spin_space_group}
                \end{equation}
The orbital-space cubic symmetry is intact in its spin space group.
The symmetry restoration results from releasing the system from the SOC constraint of Eq.~\eqref{SOC_constraint}.
The restoration is observed for many magnetic materials as well as the ferromagnet.

The macroscopic physical properties are of our interest, and thus it is enough to take into account the spin point group given by ignoring the translation operations as 
                \begin{equation}
                \spinpg (\sping) = \{ \sop{R}{W} | \sop{\left( R, \bm{t} \right)}{W} \in \sping \}.
                \label{sping_to_spinpg_for_total_sping}
                \end{equation}
According to Eq.~\eqref{spingroup_structure}, the spin point group is similarly decomposed as 
        \begin{equation}
            \spinpg (\sping) = \spinpg_\text{so} \times \overline{\spinpg}.
            \label{spin_crystallographic_point_group}
        \end{equation}
In the right-hand side, $\overline{\spinpg}$ is derived from the nontrivial spin space group $\overline{\sping}$ similarly to Eq.~\eqref{sping_to_spinpg_for_total_sping}.

Following the convention in Ref.~\cite{Litvin1977}, the spin point group is denoted by the paired operations $\spgop{W}{R}$ for $g = \sop{R}{W}$.
For example, when the spin point group is obtained such as
                \begin{equation}
                \spgop{2}{2} /\spgop{-1}{m},
                \label{spin_point_group_example1}
                \end{equation}
it is generated by a set of operations
                \begin{equation}
                \sop{2}{2},~ \sop{m}{-1}.
                \label{spin_point_group_generator_example1}
        \end{equation}
The orbital point group symmetry is given by $2/m$ whose two-fold rotation and mirror reflection are connected with the spin-space two-fold rotation and space-inversion, respectively.

Bearing in mind that the time-reversal operation is related to improper rotations in the spin space, we reduce a given spin crystallographic point group to the point groups consisting of either spin-space or orbital-space operations.
The spin-space part is defined by
                \begin{equation}
                \spog (\spinpg)= \{ W | \sop{R}{W} \in \spinpg  \},
                \label{spin_space_spinpg}
                \end{equation}
where the orbital-space operations ($R$) are irrelevant.
The orbital-space part corresponds to a well-known magnetic point group and similarly reads as
                \begin{equation}
                \mpg (\spinpg) = \{ R_W | \sop{R}{W} \in \spinpg  \},
                \label{orbital_space_spinpg}
                \end{equation}
where $R_W = R$ for det\,$W=+1$ and $R_W = R' \equiv \theta R$ for det\,$W=-1$; \textit{e.g.,} $R_W = 1' = \theta$ is the orbital time-reversal operation.
We again note that the improperness of the spin-space operation $W$ should be incorporated into the orbital-space symmetry to respect the effect of the time-reversal operation.
Similarly to the case with SOC, we refer to the orbital part as either colorless, gray, or black-white in terms of the orbital time-reversal operation $R_W = 1'$ (see Appendix~\ref{SecApp_note_group_theory}).
For the example of Eq.~\eqref{spin_point_group_example1}, the spin-space and orbital-space point groups are respectively given by
                \begin{equation}
                \spog (\spinpg) = 2/m,
                \label{spin_point_group_example1_to_spin_only_pg}
                \end{equation}
and 
                \begin{equation}
                \mpg (\spinpg) = 2/m'.
                \label{spin_point_group_example1_to_MPG}
                \end{equation}
The obtained orbital part is a black-white group.

To demonstrate the role of spin-group symmetry analysis, it is better to make a comparison to the conventional analysis based on magnetic point groups with the SOC effect.
The spin-orbital-coupled (SO-coupled) magnetic point group $\mpgsoc$ is derived from $\spinpg$ by respecting the SOC constraint of Eq.~\eqref{SOC_constraint}.
Corresponding to Eq.~\eqref{spin_point_group_generator_example1}, we obtain the colorless magnetic point group
                \begin{equation}
                \mpgsoc= 2,
                \end{equation}
where no operation involving the time-reversal operation exists in contrast to the black-white point group of Eq.~\eqref{spin_point_group_example1_to_MPG}.
The series of magnetic symmetry is summarized in Fig.~\ref{Fig_schematics}(b,c).

We aim to identify the physical phenomena emerging from the spin order without relying on SOC, and the series of point groups $(\spinpg, \spog, \mpg, \mpgsoc)$ are convenient; The orbital (spin) part of the spin point group suffices to analyze the symmetry of the object in the orbital (spin) space, while that of the SO-entangled object is determined by the overall spin point group.
For instance, recalling the transformation property of objects depicted in Fig.~\ref{Fig_symops}, each transformation in Eq.~\eqref{representation_matrices_examples} is sufficiently described by the group of Eq.~\eqref{spin_space_spinpg} for spin-space objects and of Eq.~\eqref{orbital_space_spinpg} for orbital-space objects, respectively.
On the other hand, for an example of SO-coupled operator, the spin current ($J_a^{s_b} \sim \{ p_a, s_b \}/2$) manifests the transformation property given by the direct product of representation matrices as $\hat{D}^{(p)} (g) \times \hat{D}^{(s)} (g)$.
Thus, the spin point group [Eq.~\eqref{sping_to_spinpg_for_total_sping}] is indispensable to describe the representation matrix for $J_a^{s_b}$.
In the following parts, we raise examples of spin crystallographic groups. The group symmetry is identified by the computational methods proposed in Refs.~\cite{togo2018tspglib,Shinohara2023-sy,Shinohara2024-ld}, and hence we do not show explicit derivations.

\section{Spin-group classification of physical property}
\label{Sec_classify_physical_properties}

We consider the physical properties of the magnetic materials with or without SOC on the basis of Sec.~\ref{Sec_spin_group_basic}.
Firstly, we present the spin-point-group symmetry analysis of the linear response function as well as that with the SO-coupled magnetic point group~\cite{nye1985physical}.
We classify the response into T-even and T-odd contributions, which are allowed without and with the time-reversal-symmetry breaking, respectively~\cite{Zelezny2017-ov,Watanabe2017-qk,Watanabe2018-cu,Hayami2018-bh}.
By generalizing the previous classification of dc responses~\cite{Zelezny2017-ov,Watanabe2017-qk}, we present the classification taking account of the frequency dependence (Sec.~\ref{SecSub_transport}).

In particular, we identify two aspects of the spin-group symmetry analysis through the comparative study with the magnetic point group; (1) intact symmetry in the orbital space leads to vanishing responses irrelevant to the spin degree of freedom, while it is not the case for spin-related phenomena.
(2) the nontrivial spin translation symmetry makes the spin space highly symmetric even in the presence of the spontaneous spin order and hence severely forbids spin-related phenomena such as the spin magnetoelectric effect and spin Hall effect.
These contrasting circumstances are facilitated to identify through the identification of the spin space group which can be comprised of the nontrivial spin translation group while preceding symmetry analysis is for the magnetic materials with the zero propagation vector~\cite{Zelezny2017-tf,Zhang2018-yj}.
We also introduce multipolar degrees of freedom relevant to those responses.
The identification of a given physical property is based on the developed computational method.

\subsection{Response function and T-even/T-odd decomposition}
\label{SecSub_transport}

We consider the linear response formula to illustrate the symmetry of the transport phenomena.
The formula is written by
                \begin{equation}
                X_i (\omega) = \chi_{ij}^{XY}(\omega) F_j^{(Y)} (\omega),
                \label{linear_response_formula}
                \end{equation}
where the physical quantities $X_i,Y_j$ and the force $F_j^{(Y)}$ conjugate to $Y_j$ are in the frequency ($\omega$) domain.
In the framework of the linear response theory~\cite{kubo2012statistical}, we can derive the constraint on the response coefficient from the preserved symmetry in a quantum-mechanical manner~\cite{Seemann2015-ns}.
Applying the symmetry operation $g$ of a given point group $G$, we obtain the symmetry constraint
                \begin{equation}
                \chi_{ij}^{XY} (\omega)= \chi_{kl}^{XY} (\omega) D_{ki}^{(X)} (g) D_{lj}^{(Y)} (g),
                \label{transport_unitary_constraint}
                \end{equation}
for the unitary operation and
                \begin{equation}
                \chi_{ij}^{XY} (\omega) = \chi_{kl}^{YX} (\omega) \left( D_{kj}^{(Y)} (g) \right)^\ast \left( D_{li}^{(X)} (g) \right)^\ast,
                \label{transport_anti_unitary_constraint}
                \end{equation}
for the anti-unitary operation.
We introduced the representation matrices for $X_i$ and $Y_j$ as in Eq.~\eqref{tensor_transformation}.
The anti-unitary symmetry relates the response function $\chi_{ij}^{XY}$ with $\chi_{ji}^{YX}$ for the inverse response $Y_j = \chi_{ji}^{YX} F_i^{(X)}$, in which the force $\bm{F}^{(X)}$ is required to be conjugate to $\bm{X}$.
When the current participates in the response such as the electric conductivity  $J_i = \sigma_{ij} E_j$, it is convenient to rewrite the response function by the canonical correlation function.
The symmetry argument is similarly described for the canonical correlation (see Appendix~\ref{SecApp_canonical_correlation}).

Furthermore, one can decompose the response into the \textit{symmetric} and \textit{antisymmetric} parts~\cite{Watanabe2017-qk}.
The Lehmann representation of the response function is
                \begin{align}
                \chi_{ij}^{XY} (\omega)
                        &= \sum_{ab} \frac{\rho_a - \rho_b}{\omega +i \eta + \epsilon_a - \epsilon_b} \Braket{a | X_i | b} \Braket{b | Y_j | a},\\
                        &\equiv \sum_{ab} \frac{\rho_{ab}}{\omega +i \eta + \epsilon_{ab}} X_{ab}^i Y_{ba}^j,
                \end{align}
with the adiabaticity parameter $\eta = +0$ and with $\rho_{ab} = \rho_a - \rho_b$.
The indices $a,b$ are for the eigenstates of the many-body Hamiltonian in equilibrium, $\epsilon_a$ is the eigen-energy, and $\rho_a $ is the Boltzmann factor parametrized by $\epsilon_a$.
The symmetric (s) and antisymmetric (a) parts are defined by dividing the prefactor into 
                \begin{align}
                        \frac{\rho_{ab}}{\omega +i \eta + \epsilon_{ab}} 
                                &= \frac{\omega + i\eta}{\left( \omega + i\eta \right)^2 - \epsilon_{ab}^2} \rho_{ab} - \frac{\epsilon_{ab}}{\left( \omega + i\eta \right)^2 - \epsilon_{ab}^2} \rho_{ab},\\ 
                                &= \kappa_ {ab}^\text{a} + \kappa_{ab}^\text{s}.
                        \label{reative_antisymmetric_partition}
                \end{align}
These terms show the odd or even parity under the permutation of indices $(a,b)$, and thereby we obtain the decomposition as $\chi_{ij}^{XY} = \chi_{ij}^{XY,\text{s}} + \chi_{ij}^{XY,\text{a}}$.
Similarly partitioning the product of matrix elements of $X_i$ and $Y_j$,
                \begin{align}
                        X_{ab}^i Y_{ba}^j 
                                &= \frac{1}{2}\left( X_{ab}^i Y_{ba}^j + X_{ba}^i Y_{ab}^j \right) +  \frac{1}{2}\left( X_{ab}^i Y_{ba}^j - X_{ba}^i Y_{ab}^j \right),\\
                                &\equiv \left\{ X_i, Y_j \right\}_{ab} + \left[ X_i, Y_j \right]_{ab}.
                \end{align}
After the summation over $(a,b)$, the surviving terms are $\kappa_{ab}^\text{s} \left\{ X_i, Y_j \right\}_{ab}$ and $\kappa_{ab}^\text{a} \left[ X_i, Y_j \right]_{ab}$.
It indicates that the symmetric and antisymmetric parts of the indices $(a,b)$ are respectively the symmetric and antisymmetric terms with respect to the permutation of the response and field ($X_i, Y_j$).
As a result, the symmetric and antisymmetric parts of $\chi_{ij}^{XY}$ are recast as
                \begin{align}
                        \chi_{ij}^{XY,\text{s}} (\omega) &= \frac{1}{2} \left( \chi_{ij}^{XY}(\omega) + \chi_{ji}^{YX}(\omega) \right),
                        \label{symmetric_response} \\
                        \chi_{ij}^{XY,\text{a}} (\omega)&= \frac{1}{2} \left( \chi_{ij}^{XY}(\omega) - \chi_{ji}^{YX} (\omega)\right).
                        \label{antisymmetric_response}
                \end{align}

When the time-reversal symmetry is intact, we obtain
                \begin{equation}
                        X_{ab}^i Y_{ba}^j  = \theta_X \theta_Y  X_{\bar{b}\bar{a}}^i Y_{\bar{a}\bar{b}}^j,
                \end{equation}
where $\bar{a}$ is the time-reversal partner for the eigenstate $a$ and $\theta_X$ is the parity of $X$ under the time-reversal operation.
Since the paired states have the same energy ($\epsilon_a = \epsilon_{\bar{a}}$), it is shown that only the symmetric part survives for the case of $\theta_X \theta_Y = +1$ while the antisymmetric part does when $\theta_X \theta_Y = -1$.
Once the time-reversal symmetry is lost, we obtain the other contributions, that is, the symmetric term for $\theta_X \theta_Y = -1$ and the antisymmetric for $\theta_X \theta_Y = +1$.
We label the contributions allowed in time-reversal-symmetric systems by \textit{T-even} contributions and those arising from the time-reversal-symmetry breaking by \textit{T-odd} contributions.
Note that one can utilize the orbital time-reversal symmetry $g = \sop{1}{W}$ (det$\,W = -1$) instead of the time-reversal symmetry $g = \sop{1}{-1}$, when $X$ and $Y$ are in the orbital space.
Keeping the symmetric and antisymmetric decomposition of Eqs.~\eqref{symmetric_response} and \eqref{antisymmetric_response} in mind, the T-even contribution gives rise to $\chi_{ij}^{XY,\text{s}}$ for $\theta_X \theta_Y = +1$ and $\chi_{ij}^{XY,\text{a}}$ for $\theta_X \theta_Y = -1$.
On the other hand, the T-odd contribution is complementary to the T-even, that is, $\chi_{ij}^{XY,\text{s}}$ for $\theta_X \theta_Y = -1$ and $\chi_{ij}^{XY,\text{a}}$ for $\theta_X \theta_Y = +1$.

The symmetric-antisymmetric partition and even-odd classification with respect to the time-reversal operation imply the frequency dependence of the response.
To be more specific, each term is even- or odd-order in the frequency $\omega$ as $\hat{\chi}^\text{a} \sim \omega^{2n+1}$ and $\hat{\chi}^\text{s} \sim \omega^{2n}$ in the limit of $\eta \to 0$ [Eq.~\eqref{reative_antisymmetric_partition}].
Considering the static limit ($\omega \to 0 $), two parts similarly indicate the dependence on the relaxation time $\tau$.
This can be intuitively understood by replacing the adiabaticity parameter $\eta$ with the phenomenological scattering rate as $\eta \to \tau^{-1}$.
For instance, the antisymmetric part is recast as
                \begin{equation}
                \chi_{ij}^{XY,\text{a}} \to \sum_{a,b}\frac{- i\tau^{-1}}{ \tau^{-2} + \epsilon_{ab}^2} \left[ X_i, Y_j \right]_{ab},
                \end{equation}
whose equi-energy matrix elements give rise to contribution $\sim \tau^1$ such as the Drude term of electric conductivity.
The antisymmetric term leads to the term $O(\tau^{2n+1})$ which may be characteristic of the transport phenomena in metals, and the symmetric term corresponds to the contributions as large as $O(\tau^{2n})$ including what may appear in insulators such as anomalous Hall conductivity.
In some cases, the dc antisymmetric term is labeled by an extrinsic (dissipative, absorptive) effect, while the symmetric is intrinsic (dissipation-less, reactive)~\cite{Zelezny2017-ov,Watanabe2017-qk}.

It is noteworthy that the symmetric responses are related to equilibrium properties of materials, that is physical quantities one can observe without dissipation, in some cases.
When we assume equilibrium conditions for Eq.~\eqref{linear_response_formula}, that is, the dc limit and zero antisymmetric contribution, the remaining term is solely symmetric and satisfies
                \begin{equation}
                \chi_{ij}^{XY} = \chi_{ji}^{YX},
                \label{only_symmetric_response_relation}
                \end{equation}
The symmetry of ($X_i,Y_j$) implies the phenomenological free energy given by
                \begin{equation}
                \mathcal{F}_{XY} = - \chi_{ij}^{XY} F_i^{X}F_j^{Y}.
                \end{equation}
The relation of Eq.~\eqref{only_symmetric_response_relation} is reproduced by the free energy because $X_i = - \partial \mathcal{F}_{XY} / \partial F_i^X$ and $Y_i = - \partial \mathcal{F}_{XY} / \partial F_j^Y$.
The discussion can be applied to various equilibrium properties such as piezoelectric, piezomagnetic, magnetoelectric effects, and so on~\cite{nye1985physical}.

In Table~\ref{Table_electric_magnetic_decomposition}, we summarize the classification in terms of the symmetric and antisymmetric parts.
We also list some examples of the T-even/T-odd classification by taking $(X,Y) = (\bm{J},\bm{E})$ (electric conductivity), $(\bm{M},\bm{E})$ (magnetoelectric effect~\cite{Fiebig2005-hj}, magneto-galvanic effect~\cite{Edelstein1990-jr,Ganichev2002-ek,Furukawa2017-wx}), $(\hat{\varepsilon},\bm{E})$ with strain $\varepsilon_{ij}$ (piezoelectric and magneto-piezoelectric effect), and $(\hat{\varepsilon},\bm{H})$ (piezomagnetic and kinetically-piezomagnetic effect).
The T-odd contribution for $(\hat{\varepsilon},\bm{E})$, called magnetopiezoelectric effect, has recently been proposed by theories~\cite{Varjas2016-sw,Watanabe2017-qk} and demonstrated in experiments~\cite{Shiomi2019-co,Shiomi2019-eo,Shiomi2020-ux}.

                \begin{table*}[htbp]
                \caption{Classification of the response function by symmetric and antisymmetric parts. The frequency dependence for the ac response $\hat{\chi} (\omega)$ and relaxation-time dependence for the dc response $\hat{\chi}_\text{dc}$ are listed.
                The symmetric and antisymmetric terms are further categorized by the T-even and T-odd contributions in the light of the total time-reversal parity $\theta_\text{tot} = \theta_X \theta_Y$.
                We tabulate some specific classifications for $(X,Y)$ denoted by the pair of response $X$ and field $Y$.}
                \label{Table_electric_magnetic_decomposition}
                \centering
                \begin{tabular}{ccccc}
                \toprule
                        & \multicolumn{2}{c}{antisymmetric} & \multicolumn{2}{c}{symmetric}\\
                        $\hat{\chi} (\omega)$& \multicolumn{2}{c}{$\omega^{2n+1}$} & \multicolumn{2}{c}{$\omega^{2n}$}\\
                        $\hat{\chi}_\text{dc}$& \multicolumn{2}{c}{$\tau^{2n+1}$} & \multicolumn{2}{c}{$\tau^{2n}$}\\
                \midrule
                        $\theta_\text{tot}$ & +1 & -1 & +1 & -1 \\ 
                        & T-odd & T-even & T-even & T-odd \\ 
                \midrule
                        $(\bm{X},\bm{Y})$&  & &  & \\ 
                        $(\bm{J},\bm{E})$&  & Drude &  & Hall \\ 
                        $(\bm{M},\bm{E})$&  & magnetogalvanic &  & magnetoelectric \\ 
                        $(\hat{\varepsilon},\bm{E})$& magnetopiezoelectric &  & piezoelectric &  \\ 
                        $(\hat{\varepsilon},\bm{H})$& & kinetically-piezomagnetic &  & piezomagnetic \\ 
                \bottomrule
                \end{tabular}
                \end{table*}

We introduce the symmetry of response functions based on the unitary and anti-unitary properties of symmetry operations without specifying the group.
Thus, the symmetry argument works in the case with and without SOC.
We also note that the decomposition plays a powerful role in the nonlinear response as well~\cite{Watanabe2020-oe,Watanabe2021-bt,Ahn2020-ec}.
In the same spirit of the T-even/T-odd decomposition, theoretical studies have been presented in a diagrammatic fashion~\cite{Oiwa2022-he,Michishita2022-fh}.

\subsection{Geometrical Hall effect and spin/orbital magnetization}
\label{SecSub_magnetization}

We revisit the relation between the Hall response and the magnetization from the viewpoint of the spin crystallographic group.
The Hall response reads as
                \begin{equation}
                J_i = \epsilon_{ijk} \sigma_k^\text{H} E_j,
                \end{equation}
where the Hall conductivity may be classified into three parts $\bm{\sigma}^\text{H} = \bm{\sigma}^\text{n} + \bm{\sigma}^\text{KL} + \bm{\sigma}^\text{g}$~\cite{Hirschberger2021-gk}; Normal Hall effect $\bm{\sigma}^\text{n}$ allowed under the external magnetic field, Karplus-Luttinger (KL) Hall effect $\bm{\sigma}^\text{KL}$, and geometrical (spontaneous, topological) Hall effect $\bm{\sigma}^\text{g}$.
The latter two contributions result from the magnetic ordering and are therefore summarized to the anomalous Hall effect, while they differ with respect to the role of SOC~\cite{Nagaosa2010-dn,Smejkal2022-ao}.
The KL Hall effect can appear in the presence of SOC as investigated in diverse magnetic materials such as ferromagnets, those with weak ferromagnetism, and compensated collinear and coplanar antiferromagnets~\cite{Smejkal2020-yt,Gonzalez_Betancourt2023-uq,Chen2014-fx,Higo2018-es,Wu2020-hm,Kipp2021-hv}.
The contribution may be appreciable in systems having large uniform spin magnetization ($\bm{M}_\text{sp}$) as observed that the SOC-assisted anomalous Hall conductivity is typically proportional to their uniform magnetization~\cite{Nakatsuji2022-vr}.
We note that the empirical rule is not applicable to some series of antiferromagnetic materials such as Mn$_3$Sn.
For instance, the magnetic multipolar fields offer the anomalous Hall response with the help of SOC but without the net magnetization~\cite{Suzuki2017-ps}.
In contrast, the geometrical Hall effect is characteristic of noncoplanar magnets whose geometrical texture of spins allows quasiparticles to be deflected without the help of SOC.
We identify the anomalous Hall effect driven by the spin order without SOC as the geometrical Hall effect by referring to the nontrivial geometrical texture irrespective of intrinsic or extrinsic cause, which is a noncoplanar spin structure.
The definition covers the known mechanism for the SOC-free anomalous Hall effect such as that induced by the fictitious magnetic flux arising from the spin order~\cite{Ohgushi2000-tf}.

The Hall conductivity $\bm{\sigma}^\text{H}$ is an axial and time-reversal-odd vector defined in the orbital space, coinciding with the symmetry of the orbital magnetization $\bm{M}_\text{orb}$.
Then, we can verify the anomalous Hall effect by $\bm{M}_\text{orb}$ of a given magnetic material.
Beyond the symmetry, the correlation between orbital magnetization and anomalous Hall effect can be found as clarified by the well-known St\v{r}eda formula~\cite{Streda1982-ae}.
The orbital magnetization may cover a broad range of materials hosting the anomalous Hall effect such as systems with orbital flux~\cite{Haldane1988-dm} and Graphene-based ferromagnetic systems~\cite{Polshyn2020-vp}.
The symmetry-adapted form of $\bm{M}_\text{orb}$ is computationally identified by Eq.~\eqref{tensor_transformation} with a given magnetic symmetry.
We note that the orbital magnetization exists while the localized magnetic moments are attributed to the spin degree of freedom [Eq.~\eqref{spin_order_molecular_field}] with quenched atomic orbital angular momentum.

It is of paramount interest how the KL and geometrical terms are distinguished since the distinction clarifies the SOC effect on emergent physical responses.
For instance, the geometrical effect has been intensively studied in early works; \textit{e.g.,} those with the pyrochlore ferromagnets such as Nd$_2$Mo$_2$O$_7$~\cite{Ohgushi2000-tf,Taguchi2001-kj,Shindou2001,Hirschberger2021-gk}.
Nd$_2$Mo$_2$O$_7$ undergoes the ferromagnetic order of Mo atoms and subsequently the noncoplanar magnetic order of Nd atoms as temperature decreases.
The two magnetic states with different spin-structure dimensions are labeled by the same magnetic point group in the SO-coupled case.
Two types of anomalous Hall responses therefore cannot be distinguished by the symmetry in the conventional context.
This is, however, not the case in the framework of the spin crystallographic group.

Considering the ferromagnetic Fe of Eq.~\eqref{Fe_FM_spin_space_group}, we derive the orbital part of the spin point group $\mpg$ by Eq.~\eqref{orbital_space_spinpg} to identify the symmetry of orbital magnetization $\bm{M}_\text{orb}$ dwelling on the orbital space.
The obtained $\mpg$ is a gray group written by
                \begin{equation}
                \mpg = m\bar{3}m 1',
                \end{equation}
in which the orbital time-reversal symmetry ($1'$) comes from the spin-space mirror symmetry in the spin-only group of Eq.~\eqref{spin_only_group_1dim}.
It indicates the zero orbital magnetization and vanishing anomalous Hall effect which are odd-parity under the orbital time-reversal operation, consistent with the symmetry analysis presented in Ref.~\cite{Zhang2018-yj}.
On the other hand, once the SOC is switched on, the ferroic spin magnetization is admixed with the orbital magnetization and manifests the favorable direction with respect to the crystal axes.
The resultant point group symmetry is reduced to the tetragonal magnetic point group $\mpgsoc = 4/mm'm'$ allowing for the spin-orbital-entangled magnetization along the four-fold rotation axis as derived in the established symmetry analysis~\cite{nye1985physical}.

As a result, the Hall response of Fe is attributed to the KL contribution ($\bm{\sigma}^\text{KL} \neq  \bm{0},~\bm{\sigma}^\text{g} = \bm{0}$).
This argument can be applied to spin space groups for all the one- and two-dimensional spin configurations [Eqs.~\eqref{spin_only_group_1dim},\eqref{spin_only_group_2dim}].
Then, if the magnetic moments spanning low-dimensional structure are supposed to originate from the spin magnetization, it can be said that the ordered state preserves the orbital time-reversal symmetry.
It follows that the corresponding orbital-space point group is gray in terms of the magnetic point group.
It similarly indicates the absence of physical phenomena originating from the violation of the orbital time reversal symmetry such as the orbital piezomagnetic effect (Sec.~\ref{Sec_classify_physical_properties}) and orbital magnetoelectric effect (Sec.~\ref{SecSub_geometric_magnetic_multipoles}).

                \begin{figure}[htbp]
                \centering
                \includegraphics[width=0.95\linewidth,clip]{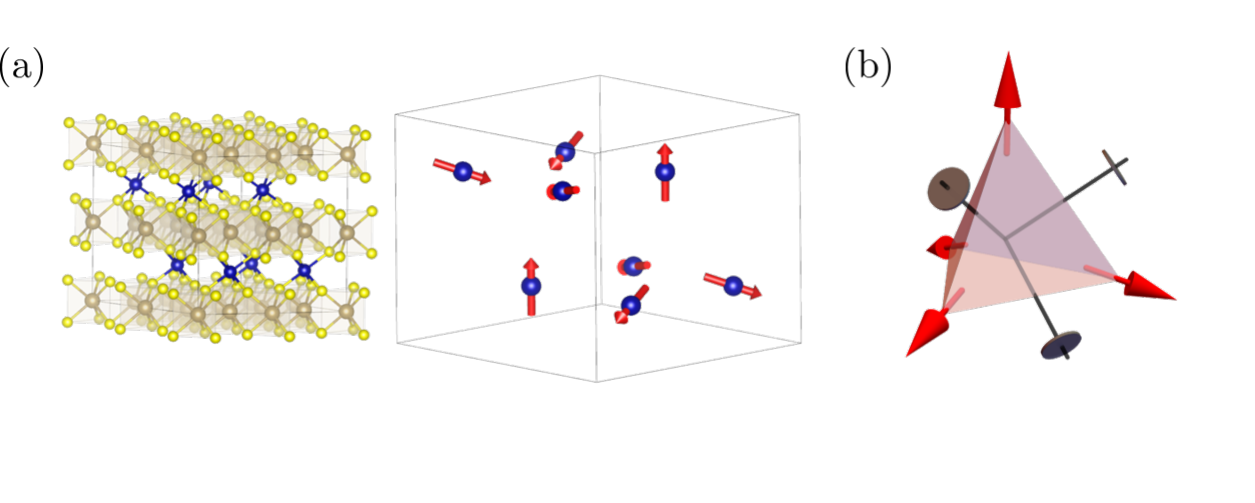}
                \caption{(a) Crystal (left panel) and spin (right panel) structures of \cota{}. (b) The tetrahedron spanned by four spins in the magnetic structure. The black lines denote the two-fold rotation axes relevant to the spin translation group.}
                \label{Fig_CoTa3S6}
                \end{figure}

The anomalous Hall response does not suffer from such a severe symmetry constraint in the case of the noncoplanar magnets because of the trivial spin-only group [Eq.~\eqref{spin_only_group_3dim}].
It is noteworthy that the nontrivial spin translation symmetry realizes the orbital magnetization not admixed with the spin counterpart. 
Let us consider a layered material \cota{} for an example~\cite{Takagi2023-ih,Park2023-zp} (Fig.~\ref{Fig_CoTa3S6}).
After the computational search for the magnetic symmetry~\cite{Shinohara2023-sy}, the magnetic space group symmetry is identified to
                \begin{equation}
                \msg = P32'.
                \label{magnetic_point_group_soc_CoTa3S6}
                \end{equation}
The associated SO-coupled magnetic point group $\mpgsoc = 32'$ leads to the conclusion that the spin ($\bm{M}_\text{sp}$) and orbital magnetization ($\bm{M}_\text{orb}$) can concurrently show up along the three-fold rotation axis.
Thus, we cannot distinguish the KL and geometrical contributions to the Hall effect within the conventional magnetic point group analysis.

Next, We consider the spin-crystallographic-group symmetry identified by the method proposed in Ref.~\cite{Shinohara2024-ld}.
Owing to the noncoplanar spin structure, the spin-only group is trivial ($\spinpg_\text{so} = \{ 1 \}$), and thereby the spin space group is coincident with the nontrivial spin space group, $\sping = \overline{\sping}$ in Eq.~\eqref{spingroup_structure}.
For the (nontrivial) spin translation group, we similarly obtain $\sping_\text{st} = \overline{\sping}_\text{st}$.
The spin space group is written by the internal semidirect product of the spin translation group $\sping_\text{st}$ and the remaining part $\mathcal{H}$;
                \begin{equation}
                \sping = \sping_\text{st} \rtimes \mathcal{H}.
                \label{spin_space_group_CoTa3S6}
                \end{equation}
Since we are interested in the macroscopic physical property, it is sufficient to take into account the point group symmetry.
The point group is obtained from Eq.~\eqref{spin_space_group_CoTa3S6} as
                \begin{equation}
                \spinpg = \spinpg_\text{st} \rtimes \spinpg_{\mathcal{H}}.
                \end{equation}
The point groups $\spinpg_\text{st}$ and $\spinpg_{\mathcal{H}}$ are respectively derived from $\sping_\text{st}$ and $\mathcal{H}$ as in Eq.~\eqref{sping_to_spinpg_for_total_sping}.
The spin-translation part $\spinpg_\text{st}$ gives rise to the spin-only-group symmetry written by
                \begin{equation}
                \spinpg_\text{st} = \left\{ \sop{1}{W} |  W \in \{ 1, 2_X, 2_Y, 2_Z \} \right\}.
                \label{spin_translation_symmetry_CoTa3S6}
                \end{equation}
The point group $222 = \{ 1, 2_X, 2_Y, 2_Z \} $ is given by the mutually-orthogonal two-fold rotation axes $(X,Y,Z)$ by which four orientations of Co spins are interchanged [Fig.~\ref{Fig_CoTa3S6}(b)].
The remaining part is
                \begin{equation}
                \spinpg_{\mathcal{H}} = \spgop{3}{6}~ \spgop{m_{(100)}}{2} ~ \spgop{m_{(010)}}{2}. 
                \label{nontrivial_spin_point_group_CoTa3S6}
                \end{equation}
We notice that the six-fold symmetry of the crystal (space group $P6_3 22$, No.~182) is intact in the ordered phase without the SOC effect.

We are interested in the spin and orbital magnetization which are related to the KL and geometrical Hall effects, respectively.
Then, it is enough to consider the spin and orbital parts projected from $\spinpg$ as in Eqs.~\eqref{spin_space_spinpg} and \eqref{orbital_space_spinpg}.
The spin part is 
                \begin{equation}
                \spog (\spinpg) = 222 \rtimes 3m = \bar{4}3m,
                \label{spinpg_to_spinonly_group_CoTa3S6}
                \end{equation}
and the orbital part is
                \begin{equation}
                \mpg (\spinpg) = 62'2'.
                \label{spinpg_to_orbitalonly_group_CoTa3S6}
                \end{equation}
The spin translation symmetry of Eq.~\eqref{spin_translation_symmetry_CoTa3S6} leads to the cubic symmetry $\spog = \bar{4}3m$ despite the hexagonal crystal structure of \cota{}.
The spin point group symmetry enhanced by the spin translation symmetry has not been addressed in previous studies of emergent responses.
On the other hand, the orbital part manifests the axial symmetry whose rotation axis is $[001]$ similar to the SO-coupled case of Eq.~\eqref{magnetic_point_group_soc_CoTa3S6}.
Using the spin and orbital parts in the spin point group, we identify the allowed spin and orbital magnetization,
                \begin{equation}
                \bm{M}_\text{sp} = \bm{0},~\bm{M}_\text{orb} \parallel [001].
                \label{magnetization_CoTa3S6}
                \end{equation}
As a result, the anomalous Hall conductivity vector $\bm{\sigma}^\text{H} \parallel [001]$ can appear without the help of SOC ($\bm{\sigma}^\text{g} \neq \bm{0}$).
Furthermore, the zero spin magnetization follows from the cubic symmetry in the spin space.
These properties indicate that the anomalous Hall effect of \cota{} can be attributed to the geometrical Hall effect ($\bm{\sigma}^\text{g}$) and that the KL contribution ($\bm{\sigma}^\text{KL}$) may be suppressed due to vanishing spin polarization since it is empirically expected to be proportional to the uniform magnetization.

We have clarified two characteristic aspects of spin group symmetry in this section.
The spin-only group associated with low-dimensional spin configuration gives rise to strong constraints on the orbital-space objects, forbidding physical responses arising from the violation of the orbital time-reversal symmetry.
On the other hand, the noncoplanar spin structure may lead to such emergent responses from orbital degrees of freedom, while the responses relevant to the spin degree of freedom may vanish due to the high symmetry of the spin space originating from the spin translation group.
These contrasting situations are systematically understood by the spin-crystallographic-group symmetry analysis incorporating more details of spin structures such as the spin-structure dimension $\spdim$ and spin-translation symmetry beyond the magnetic space group.

\subsection{Magnetoelectric effect and spin/orbital magnetic quadrupole moments}
\label{SecSub_geometric_magnetic_multipoles}

The spin-crystallographic-group symmetry analysis distinguishes the role of spin and orbital degrees of freedom in the magnetoelectric effect because it separately identifies the spin and orbital magnetization as in Eq.~\eqref{magnetization_CoTa3S6}.
We here consider the correlation between magnetization and electric polarization written by
                \begin{equation}
                P_i = \chi_{ij}^{PM} H_j, ~ M_i = \chi_{ij}^{MP} E_j.
                \end{equation}
The frequency dependence is suppressed.
In particular, the symmetric term is called magnetoelectric effect and the antisymmetric is the inverse magneto-galvanic effect~\cite{Ganichev2002-ek,Fiebig2005-hj} (see also Table~\ref{Table_electric_magnetic_decomposition}).
In the following, we focus on the DC magnetoelectric effect $\alpha_{ij} = \chi_{ij}^{MP,\text{s}} (\omega =0)$, which appears even in systems with no electric conductivity, that is, insulators at the zero temperature.
The magnetoelectric effect is further divided into the spin and orbital parts as 
                \begin{equation}
                        \alpha_{ij} = \alpha_{ij}^\text{sp} + \alpha_{ij}^\text{orb}.
                \end{equation}
where the spin and orbital magnetization participate in the response, respectively.
In the light of the spin crystallographic group, the symmetry of spin and orbital magnetoelectric effect is respectively determined by the whole of and orbital part of the spin point group, since the former is a spin-orbital-coupled response and the latter consists of only the orbital degree of freedom.

\begin{figure}[htbp]
                \centering
                \includegraphics[width=0.88\linewidth,clip]{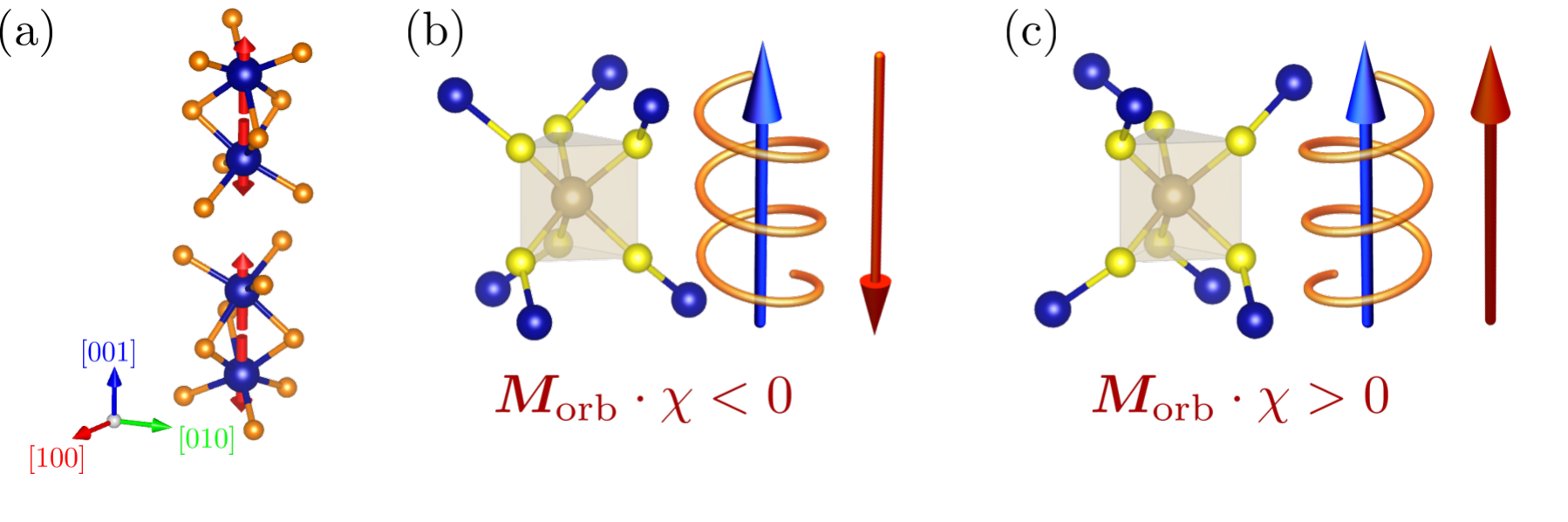}
                \caption{        
                (a) Crystal and spin structures of Cr$_2$O$_3$ where spins are collinear along the $[001]$ direction.
                (b,c) Chirality of the crystal structure of \cota{} implied by TaS$_6$ prism with the twisted coordination of Co atoms.
                The twisting arrangement determines the chirality $\chi = \pm 1$.
                When orbital magnetization (blue-colored vector) is formed along the $[001]$ direction due to its noncoplanar spin ordering, (b) the toroidal moment (red-colored vector) is anti-parallel to it, (c) while it is parallel with the opposite chirality.
                }
                \label{Fig_magnetoelectric}
                \end{figure}

Firstly, we consider a prototypical magnetoelectric material Cr$_2$O$_3$~\cite{Astrov1961-nd,Folen1961-mo} [Fig.~\ref{Fig_magnetoelectric}(a)].
Its collinear antiferromagnetic order does not break translation symmetry due to the zero propagation vector of the spin configuration, and the nontrivial spin translation group is equal to the translation group for the paramagnetic state as $\overline{\sping}_\text{st} = \mathcal{T}$.
Supposing that the spins are aligned to the $[001]$ axis in the spin space, the spin crystallographic point group is
                \begin{equation}
                \spinpg = \spinpg_\text{so} \times \overline{\spinpg},
                \label{Cr2O3_spinpg}
        \end{equation}
where the spin-only group $\spinpg_\text{so}$ is for the one-dimensional spin configuration [Eq.~\eqref{spin_only_group_1dim}]
                \begin{equation}
                        \spinpg_\text{so} = \text{SO(2)} \rtimes \left\{ 1, m_\parallel \right\}.
                        \label{Cr2O3_spin_only_group}
                \end{equation}
The remaining part is
                \begin{equation}
                \overline{\spinpg} = \spgop{\bar{3}}{\bar{3}}\spgop{m}{m}.
                \end{equation}
The spin part associated with $\spinpg$ manifests centrosymmetric point group symmetry as
                \begin{equation}
                \spog ( \spinpg) =\text{O(2)} \rtimes \{ 1, m_\parallel \},
                \label{Cr2O3_spinpg_to_spin_only}
                \end{equation}
and the orbital part is given by a centrosymmetric gray group
                \begin{equation}
                \mpg (\spinpg) = \bar{3}m 1',
                \label{Cr2O3_spinpg_to_orbital_only}
                \end{equation}
which differs from the noncentrosymmetric black-white point group $\mpgsoc = \bar{3}' m'$ for the SO-coupled case.

The absence of orbital contribution ($\alpha_{ij}^\text{orb} = 0$) follows from either of the space-inversion or time-reversal symmetry in the orbital part of the spin point group [Eq.~\eqref{Cr2O3_spinpg_to_orbital_only}] because of the odd parity under those operations~\cite{Fiebig2005-hj}.
On the other hand, when taking into account the spin-space proper rotations, the spin point group of Eq.~\eqref{Cr2O3_spinpg} does not preserve the time-reversal or space-inversion symmetry as $\spgop{-1}{1}, \spgop{1}{-1} \not\in \spinpg$ and may allow for finite spin magnetoelectric effect.
The symmetry of the spin contribution is obtained as follows.
In the SO-coupled case with the magnetic point group $\mpgsoc = \bar{3}' m'$, the allowed SO-entangled magnetoelectric effect is given by all the diagonal components $\alpha_{xx},\alpha_{yy},\alpha_{zz}$~\cite{nye1985physical}.
In the absence of SOC, one should consider additional constraints due to the spin-only group [Eq.~\eqref{Cr2O3_spin_only_group}].
The SO$(2)$ symmetry in the spin-only group forbids the spin polarization response transverse to the collinear axis ($\alpha_{yj}^\text{sp},\alpha_{zj}^\text{sp} = 0$) but allows the longitudinal as $\alpha_{zj}^\text{sp} \neq  0$.
The symmetry analysis is summarized as
                \begin{equation}
                \alpha_{zz}^\text{sp} \neq 0 \text{ otherwise } \alpha_{ij}^\text{sp} = 0,~\alpha_{ij}^\text{orb} = 0.
                \label{Cr2O3_magnetoelectricity}
                \end{equation}
As a result, only the longitudinal magnetization can respond to the applied electric field in a SOC-free manner and is purely ascribed to the spin origin.

Although we assumed the DC case, the symmetry analysis similarly holds for the AC responses.
The AC magnetoelectric effect denoted by $\chi_{ij}^{MP,\text{h}}(\omega)$ triggers the nonreciprocal optical activity~\cite{Hornreich1967-bo,Hayashida2022-xk,Malashevich2010-uw}.
Owing to the effective space-inversion ($\spgop{W}{\bar{1}} $ with det\,$W=1$) and orbital time-reversal symmetry ($\spgop{W}{1}$ with det\,$W=-1$) in the spin point group of Eq.~\eqref{Cr2O3_spinpg}, the optical activity arises solely from the spin magnetic-dipole transition but does not include the orbital magnetic-dipole or electric-quadrupole effects in the absence of SOC.

Next, we again consider \cota{} to demonstrate the role of spin translation symmetry in the magnetoelectric effect.
Similarly to zero spin magnetization, the cubic spin-space symmetry [Eq.~\eqref{spinpg_to_spinonly_group_CoTa3S6}] forbids the spin magnetoelectric effect, $\alpha_{ij}^\text{sp}= 0 $.
On the other hand, the orbital part of Eq.~\eqref{spinpg_to_orbitalonly_group_CoTa3S6} leads to finite orbital magnetoelectric effects $\alpha_{xy}^\text{orb} = -\alpha_{yx}^\text{orb}$.
Then, the magnetoelectric effect of \cota{} is summarized as
                \begin{equation}
                \alpha_{ij}^\text{sp}= 0,~\alpha_{xy}^\text{orb} = -\alpha_{yx}^\text{orb}\neq 0.
                \label{orbital_magnetoelectric_effect_CoTa3S6}
                \end{equation}
In the SO-coupled point group symmetry of Eq.~\eqref{magnetic_point_group_soc_CoTa3S6}, $\alpha_{xy} = -\alpha_{yx}$ is similarly allowed while the spin effect is admixed.

The symmetry analysis shows the possibility of the SOC-free magnetoelectric effect dominated by the orbital contribution.
Among known mechanisms for the magnetoelectricity~\cite{Tokura2014-ix}, the identified response may originate from the exchange striction mechanism~\cite{Sergienko2006-is,Solovyev2008-tx} and the dynamical phase~\cite{Mostovoy2011-gq} which do not require SOC.
Note that we here discussed the orbital magnetoelectric effect induced by the noncoplanar spin order~\cite{Delaney2009-od} rather than that what arises from the orbital-current order~\cite{Bulaevskii2008-bb}.

We took the overview of the relation between the anomalous Hall effect and orbital magnetization in Sec.~\ref{SecSub_magnetization}.
A similar discussion can be found in the case of the magnetoelectricity; the response may be correlated with higher-order anisotropy of magnetic charge, that is, the magnetic quadrupole moment $\mq_{ij}$~\cite{Spaldin2008-zj,Thole2016-zk}.
The symmetry of $\mq_{ij}$ is schematically given by the tensor product of the magnetization and position as $\mq_{ij} \sim M_i r_j$.
According to the space-time symmetry, we can find the correspondence between the magnetic quadrupole moments and the magnetoelectric effect given by
                \begin{equation}
                \mq_{ij} \leftrightarrow \alpha_{ij}.
                \label{magnetoelectric_quadrupole_relation}
                \end{equation}
The magnetization $M_i$ can be classified into the spin and orbital contributions in terms of the spin crystallographic group.
Then, the symmetry analysis of the allowed magnetoelectric effect can be reproduced by identifying the relevant spin/orbital magnetic quadrupole moments with the use of Eq.~\eqref{tensor_transformation}.

For instance, let us consider the multipolar degree of freedom corresponding to the magnetoelectric effect of \cota{}.
The allowed multipole moment is the orbital toroidal moment $\mq_{xy}^\text{orb} - \mq_{yx}^\text{orb} \sim \left( \bm{M}_\text{orb} \times \bm{r} \right)_z$ where the toroidal moment is a time-reversal-odd polar vector.
The symmetry of the toroidal moment is consistent with that of the orbital magnetoelectric effect in Eq.~\eqref{orbital_magnetoelectric_effect_CoTa3S6}.
The toroidal moment polarized along the $[001]$ direction can be intuitively understood by its orbital magnetization and crystal structure.
The space group symmetry of \cota{} (No.~182, $P6_322$) does not have any improper rotation symmetry in the orbital space~\cite{Inui2020-eq}, and hence every axially symmetrical quantity can be coupled to the corresponding polar-symmetry quantity with preserving the time-reversal parity.
In the present case, the orbital magnetization (time-reversal-odd axial vector) is coupled to the orbital toroidal moment ( time-reversal-odd vector) as in the case of magnetochiral anisotropy~\cite{Rikken2005-sp} [Fig.~\ref{Fig_magnetoelectric}(b,c)].
In the present case, the orbital magnetization (time-reversal-odd axial vector) is coupled to the orbital toroidal moment (time-reversal-odd polar vector) as in the case of magnetochiral anisotropy~\cite{Rikken2005-sp} [Fig.~\ref{Fig_magnetoelectric}(b,c)].
The coupling between the orbital magnetization and toroidal moment is a consequence of its chiral crystal structure, and the ligands surrounding magnetic Co atoms play essential roles.
We checked that the noncoplanar spin structure of Co atoms does give rise to orbital magnetization but manifests no orbital toroidal moment without Ta and S atoms.
Interestingly, beyond the symmetry analysis, recent theoretical studies identified that the magnetic quadrupole moment~\cite{Gao2018-xt,Shitade2018-vh,Gao2018-gl,Bhowal2022-ay} covers not only the magnetoelectric effect but also other cross-correlated responses~\cite{Shitade2019-it} when the system is insulating.

\subsection{Spin Hall response and rotators}

Let us consider another spin-related response, the electric-field ($E_j$) induction of spin-polarized current $J_i^{s_k}$.
The spin-polarized current may be given by $J_i^{s_k} = \{ J_i, s_k\}/2$ which is comprised of the spin- and orbital-space objects.
The response formula is given by
        \begin{equation}
        J_i^{s_k} = \sigma_{ij}^k E_j.
        \label{spin_polarized_current_response_to_E}
        \end{equation}
The parity under the time-reversal symmetry is $\theta_{J^s} \theta_E = +1$, and the T-even contribution is symmetric while the T-odd is antisymmetric according to the classification in Sec.~\ref{SecSub_transport}.
For the Hall response ($\epsilon_{ijp} \sigma_{ij}^k$) in the DC limit, the T-even effect includes the well-known spin Hall effect prominent in the spin-orbit-coupled semiconductors~\cite{Murakami2003-mq,Sinova2004-dt}, while the T-odd called the magnetic spin Hall effect is unique to magnetic metals~\cite{Freimuth2014-zz,Zelezny2017-tf,Zhang2018-yj,Naka2019-pf,Kimata2019-xe,Karube2022-et}.
The absence of SOC and spin order, indicating the isotropic spin-space symmetry, leads to the vanishing response.

\begin{figure}[htbp]
                \centering
                \includegraphics[width=0.90\linewidth,clip]{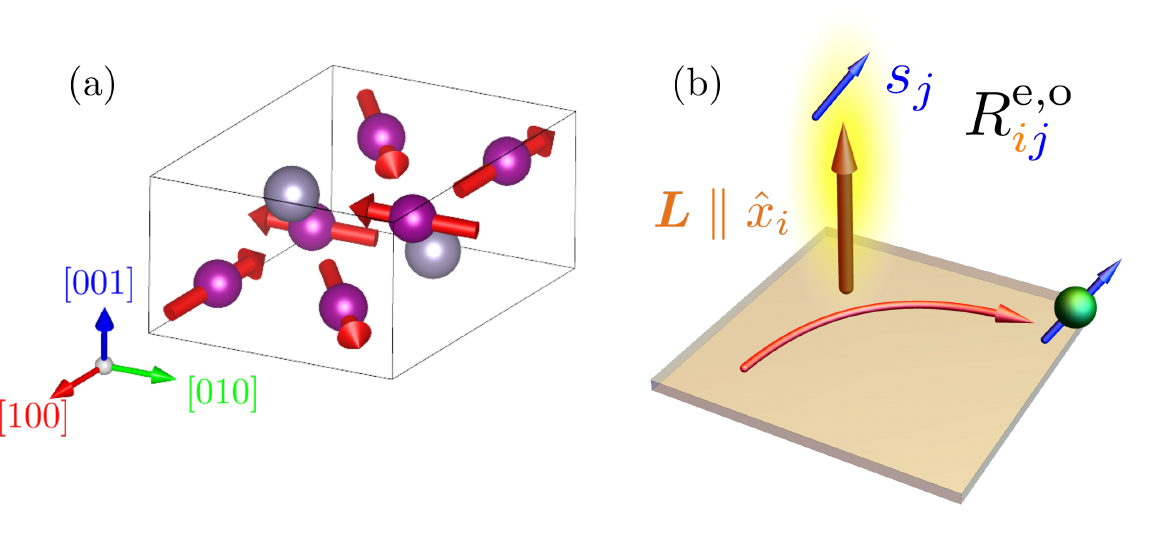}
                \caption{(a) Spin structure of Mn$_3$Sn. (b) Transverse conversion between the charge and spin currents denoted by the rotator $R_{ij}$. The Hall plane perpendicular to the $x_i$ direction is for the current whose spin is polarized along the $x_j$ direction in the spin space.}
                \label{Fig_Mn3Sn_and_rotators}
                \end{figure}

We refer to the symmetry analysis of Refs.~\cite{Zelezny2017-tf,Zhang2018-yj} and decompose $\sigma_{ij}^k$ into the T-even and T-odd components.
The target material, a noncollinear but coplanar magnet Mn$_3$Sn, is attracting a lot of attention because of its potential application for spintronic and magneto-optical components [Fig.~\ref{Fig_Mn3Sn_and_rotators}(a)]~\cite{Nakatsuji2022-vr,Smejkal2022-ao}.
The spin crystallographic point group is given by
                \begin{equation}
                \spinpg = \spinpg_\text{so} \times \overline{\spinpg}.
                \label{Mn3Sn_spinpg_total}
                \end{equation}
The spin-only group is
                \begin{equation}
                \spinpg_\text{so} =\left\{ \sop{1}{W} | W \in \{ 1, m_{(001)}\}\right\}, 
                \end{equation}
for a two-dimensional spin structure.
The remaining part is~\cite{Liu2022-dh}
                \begin{equation}
                        \overline{\spinpg} = \spgop{3}{6}/\spgop{1}{m}\spgop{m_{(120)}}{m}\spgop{m_{(110)}}{m},
                \end{equation}
where the spin configuration is taken to preserve the spin point group symmetry for $g = \sop{2_{[100]}}{2_{[100]}}$.
The spin part associated with $\spinpg$ is 
                \begin{equation}
                \spog (\spinpg) = m_{(001)} \times 3m = \bar{6}2m,
                \end{equation}
and the orbital part is
                \begin{equation}
                \mpg (\spinpg) = 6/mmm1',
                \end{equation}
coinciding with the SO-coupled magnetic point group for the paramagnetic state, while the SO-coupled magnetic point group for the magnetic state shows the crystal-class reduction from hexagonal to orthorhombic as $\mpgsoc = mm'm'$.
Thus, as far as only the orbital degrees of freedom are concerned, the antiferromagnetic state of Mn$_3$Sn does not show any physical phenomena arising from the symmetry breaking.
On the other hand, owing to the spontaneously-emerged anisotropy in the spin space, the electric field can stimulate the spin-polarized current.
The response is explicitly given by
                \begin{equation}
                \sigma_{xy}^{z} = - \sigma_{yx}^{z},
                \label{spin_polarized_current_electric_Mn3Sn}
                \end{equation}
for the T-even contributions and 
                \begin{equation}
                \sigma_{xx}^{x} = \sigma_{xy}^{y} = \sigma_{yx}^{y}= - \sigma_{yy}^{x},
                \label{spin_polarized_current_magnetic_Mn3Sn}
                \end{equation}
for the T-odd components.
These spin-orbit-free components may overwhelm those requiring the SOC effect~\cite{Zelezny2017-tf,Zhang2018-yj}.

Following the discussion parallel to those in the previous sections~\ref{SecSub_magnetization} and \ref{SecSub_geometric_magnetic_multipoles}, the symmetry analysis is applicable to more complex spin structures and can be extended to cover the orbital counterpart such as the orbital-current Hall effect, which is denoted with the current whose magnetic polarization is attributed to the orbital origin~\cite{Bernevig2005-hx,Kontani2009,Go2023-yi}.
For instance, CoTa$_3$S$_6$ with the cubic spin-space symmetry [Eq.~\eqref{spinpg_to_spinonly_group_CoTa3S6}] leads to the zero spin-current response $\sigma_{ij}^{k,\text{sp}} = 0$.
The orbital counterpart, however, is allowed even without SOC such as the T-odd contributions $\sigma_{xx}^{z,\text{orb}},\sigma_{yy}^{z,\text{orb}}$, and $\sigma_{zz}^{z,\text{orb}}$.

We further consider the quantities relevant to the T-even/T-odd spin Hall responses by the analogy of the SOC Hamiltonian~\cite{Hayami2018-bh}.
Recalling the expression for the atomic SOC of Eq.~\eqref{SOChamiltonian}, one can replace the orbital angular momentum $\bm{L}\sim \bm{r} \times \bm{p}$ with the cross product of the electric field and current $\bm{E} \times \bm{J}$, since the space-time symmetry is same in pairs of vectors ($\bm{r},\bm{E}$) and ($\bm{p},\bm{J}$).
Then, the relativistic spin-orbit interaction may correspond to the spin Hall response as  
                \begin{equation}
                \bm{L} \cdot \bm{s} \sim \left(  \bm{J} \times \bm{E} \right)\cdot \bm{s} = \epsilon_{ijk} E_j J_i s_k \leftrightarrow \sigma_{ij}^k = \epsilon_{ijk} \sigma_0,
                \label{SOC_and_spinHall}
                \end{equation}
where $\epsilon_{ijk} \sigma_0$ indicates the spin Hall effect whose spin polarization is perpendicular to the Hall plane defined by the $\bm{E}$ and $\bm{J}$ [Fig.~\ref{Fig_Mn3Sn_and_rotators}(b)].
That is why the spin Hall effect generically exists under the SOC effect.
The correspondence between the T-even spin Hall effect and the product of $L_i$ and $s_j$ may be generalized to that in the framework of the spin crystallographic group.
Then, we here introduce the \textit{T-even rotator} $R_{ij}^\text{e}$ giving the transverse correlation between charge and spin currents.
The symmetry of $R_{ij}^\text{e}$ coincides with the product of the time-reversal-odd axial vectors $L_i$ and $s_j$ which are defined in the orbital and spin space, respectively.
The T-even rotator may be attributed to the spin-resolved Berry curvature playing a crucial role in the intrinsic spin Hall effect.
The T-even rotator denotes the Hall response for the $x_j$-polarized spin current denoted by the Hall plane perpendicular to the $i$-direction [Fig.~\ref{Fig_Mn3Sn_and_rotators}(b)],
                \begin{equation}
                R_{ij}^\text{e} \leftrightarrow \epsilon_{iab} \sigma_{ab}^j
                \end{equation}
Specifically, the trace $\sum_i R_{ii}^\text{e}$ corresponds to Eq.~\eqref{SOChamiltonian}.
For instance, referring to the spin crystallographic point group of Mn$_3$Sn [Eq.~\eqref{Mn3Sn_spinpg_total}], we identify the T-even rotator $R_{zz}^\text{e} \neq 0 $ corresponding to the spin Hall response of Eq.~\eqref{spin_polarized_current_electric_Mn3Sn} where the $z = (001)$ Hall plane is obtained as $(\bm{E} \times \bm{J})_z$ and the spin current is polarized along the $z$-direction.
The T-even spin Hall effect is a response characteristic of noncollinear spin systems under no SOC effect~\cite{Zhang2018-yj} as we corroborate in Sec.~\ref{Sec_magndata_study}.

It is straightforward to derive the similar quantity relevant to the magnetic spin Hall effect, that is, T-odd rotator $R_{ij}^\text{o} \sim \overline{L}_i s_j$ with the time-reversal-even axial vector $\overline{L}_i$ defined in the orbital space~\cite{Hayami2018-bh}.
The symmetry of the T-odd rotator agrees with the spin-current vorticity clarified in a recent theoretical study~\cite{Mook2020-ae}.
In the case of Mn$_3$Sn, the T-odd rotator is absent without the SOC effect ($R_{ij}^\text{o} = 0$).
This is consistent with the symmetry analysis of Eq.~\eqref{spin_polarized_current_magnetic_Mn3Sn} whose field and response can be longitudinal to each other. 

The symmetry analysis based on rotators can be applied to spin-current responses to another stimulus such as the temperature gradient, as long as the replaced field shares the same symmetry as the electric field, that is a time-reversal-odd polar vector.
Then, the high-throughput symmetry analysis presented in the following section allows us to identify SOC-free spincaloritronic responses such as anomalous spin Nernst effect.

\section{High-throughput symmetry analysis of spin group symmetry}
\label{Sec_magndata_study}

The computational search for the spin space group allows us to identify physical properties free from the SOC effect~\cite{Shinohara2023-sy,Shinohara2024-ld}.
We present symmetry analysis with dozens of observed spin configurations obtained from \textsc{magndata}~\cite{Gallego2016-pb,Gallego2016-yz}.
We have performed the symmetry analysis of 1512 magnetic materials which have no site disorder.
For the spin-structure dimension, 914 collinear, 403 coplanar, and 195 noncoplanar spin systems are studied.
The magnetic materials are numbered by following the identification number provided in \textsc{magndata} such as Cr$_2$O$_3$ (\#\,0.59).

In this section, we discuss the physical quantities such as spin/orbital magnetization and magnetic quadrupole moments, and T-even/T-odd rotators introduced in Sec.~\ref{Sec_classify_physical_properties} to investigate emergent physical phenomena.
Providing some examples of spin space group ($\sping$) with comparison to the analysis with magnetic space group ($\msg$), we investigate characteristic physical properties in the viewpoint of symmetry.
Although the electromagnetic responses relying on nonrelativistic spin-charge coupling has been mainly discussed for spin structures with the zero propagation vector, our high-throughput symmetry analysis further identifies candidate materials offering intriguing physical phenomena arising from a complex spin structure such as purely-orbital magnetoelectric effect and motivates us to revisit known materials from the perspective of SOC-free responses.

\subsection{Spin crystallographic symmetry}

\begin{figure}[htbp]
                \centering
                \includegraphics[width=0.50\linewidth,clip]{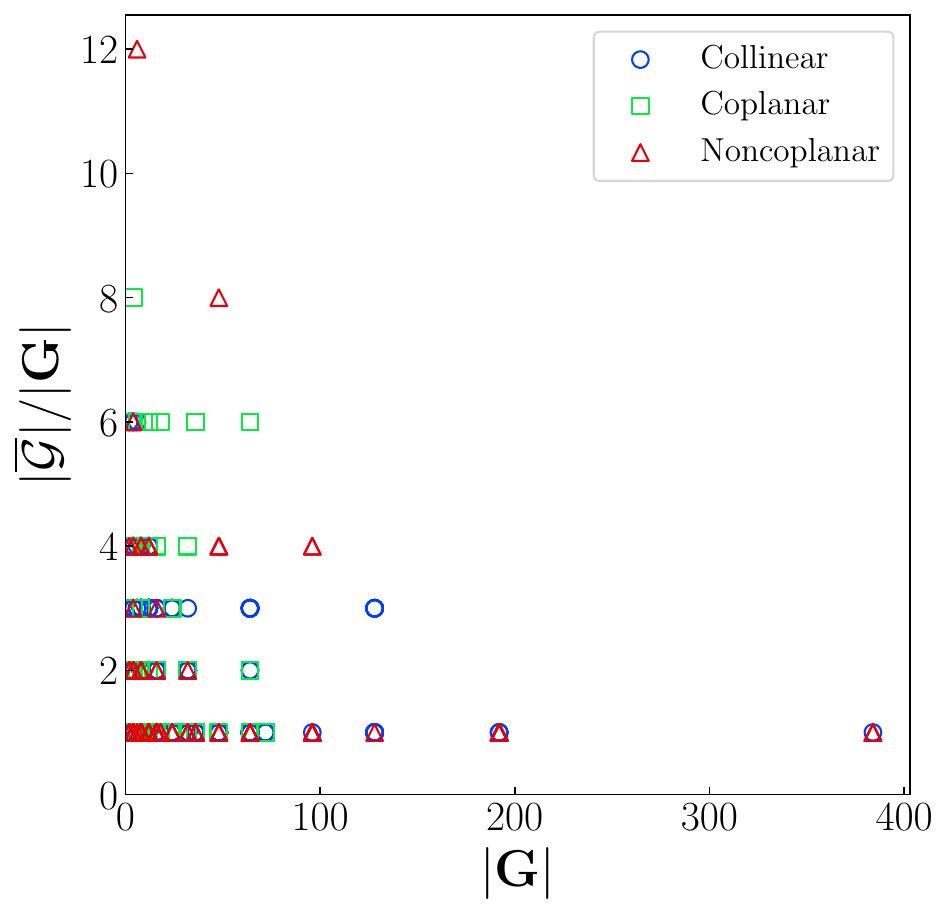}
                \caption{Distribution of magnetic materials parametrized by the order of the magnetic space group ($|\msg|$) and by the ratio between the order of the nontrivial spin space group ($|\overline{\sping}|$) and $|\msg|$.
                The distributions are categorized based on whether the spin-structure dimension is collinear, coplanar (but noncollinear), or noncoplanar.}
                \label{Fig_num_operations}      
                \end{figure}

Let us classify the magnetic materials in terms of the spin crystallographic or magnetic space group symmetry.
Since the spin space group comprises its corresponding magnetic space group as a subgroup, the orders of groups satisfy the relation $|\overline{\sping}|/|\msg| \in \mathbb{N} = \{1,2,3,\cdots \}$ where we consider the nontrivial spin space group $\overline{\sping}$ instead of $\sping$.
Figure~\ref{Fig_num_operations} illustrates how many symmetry operations are restored by neglecting SOC.
For instance, the maximal symmetry restoration occurs in the case of a noncoplanar magnet CrSe (\#\,2.35)~\cite{Corliss1961_CrSe}.
The hexagonal crystalline symmetry (space group No.~194, $P6_3/mmc$) is intact for the nontrivial spin space group $\overline{\sping}$, while the crystal class is reduced to the trigonal for the magnetic space group $\msg$. 
The nontrivial spin translation group as large as $|\overline{\sping}_\text{st}| = 3$ also contributes to the higher symmetry of $\overline{\sping}$~\cite{Shinohara2024-ld}.

For a more detailed comparison, we classify the orbital symmetry of the spin point group ($\mpg$) and the SO-coupled magnetic point group ($\mpgsoc$) in terms of the time-reversal symmetry such as colorless, gray, and black-white groups (see Appendix~\ref{SecApp_note_group_theory}).
Owing to the spin-only group, the low-dimensional spin structure ($\spdim =1,2$) makes the orbital-space symmetry gray irrespective of its SO-coupled magnetic point group symmetry [see examples of Eq.~\eqref{Fe_FM_spin_space_group} for $\spdim=1$ and Eq.~\eqref{Mn3Sn_spinpg_total} for $\spdim=2$].
On the other hand, the noncoplanar system ($\spdim=3$) can be characterized by any of three different types of magnetic point groups.
Table~\ref{Table_MPG_classification} shows the classification result.
The absence of the SOC constraint [Eq.~\eqref{SOC_constraint}] allows for the additional symmetry related to the time-reversal operation and hence some of the colorless groups among $\mpgsoc$ are turned into black-white with respect to $\mpg$.

                \begin{table}[htbp]
                \caption{Classification table of the SO-coupled magnetic point group ($\mpgsoc$) and the orbital part of the spin point group ($\mpg$) for the noncoplanar magnets in light of the time-reversal symmetry. Note that all the $\mpg$ is gray for the collinear and coplanar magnets irrespective of the type of $\mpgsoc$.  
                }
                \label{Table_MPG_classification}
                \centering
                \setlength{\tabcolsep}{15pt}
                \begin{tabular}{lccc}
                 \toprule
                   & Colorless & Gray& Black-White \\
                 \midrule
                 $\mpgsoc$&46 & 46 & 103\\
                 $\mpg$&42 & 46 & 107\\
                 \bottomrule
                \end{tabular}
                \end{table}

\begin{figure}[htbp]
                \centering
                \includegraphics[width=0.95\linewidth,clip]{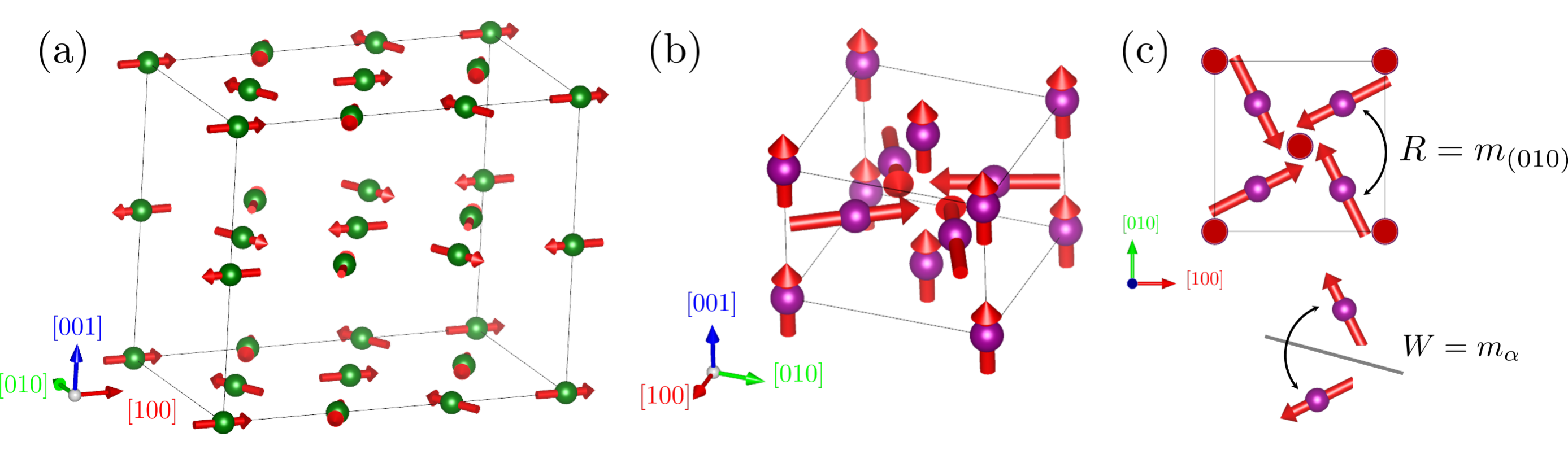}
                \caption{Spin configurations of (a) Ba$_3$MnSb$_2$O$_9$ with only Mn atoms and of (b,c) Mn$_3$CuN.
                In (c), the spin space operation $\sop{(m_{(010)},\bm{0})}{m_{\alpha}}$ is depicted.
                The mirror operations $W = m_{\beta},m_{\gamma}$ can be similarly obtained.}
                \label{Fig_spinsg_examples}
                \end{figure}

We take some examples to compare the magnetic symmetry with and without SOC.
For an example of low-dimensional spin structures, we consider a coplanar magnet Ba$_3$MnSb$_2$O$_9$ (\#\,1.0.46).
The material crystallizes in the structure denoted by the centrosymmetric space group $C2/m$ (No.~15)~\cite{Doi_2004}, and its coplanar spin structure ($\spdim = 2$) is formed by magnetic moments at Mn atoms [Fig.~\ref{Fig_spinsg_examples}(a)]. 
No symmetry including the time-reversal operation exists in the magnetic space group $\msg  = C2$ (type I) allowing for both of ferroelectric and ferromagnetic polarizations.
On the other hand, such multiferroic property is missing if there exists no SOC effect.
The spin space group reads as
                \begin{equation}
                \sping = \sping_\text{so} \times \overline{\sping}.
                \end{equation}
The two-dimensional spin structure corresponds to the spin-only group
                \begin{equation}
                \sping_\text{so} = \{ \sop{(1,\bm{0})}{1},~\sop{(1,\bm{0})}{m_{(001)}}\},
                \end{equation}
and the nontrivial part $\overline{\sping}$ is isomorphic to the space group $C2/m$.
$\overline{\sping}$ is generated by
                \begin{equation}
                        \sop{(1,\bm{t})}{1}, \sop{(-1,\bm{0})}{m_{(010)}},~\sop{(2_{[010]},\tilde{\bm{t}})}{m_{(100)}},
                \end{equation}
in addition to the trivial translation operations associated with the monoclinic crystal structure.
We here introduced the translations $\bm{t} = (0.5,0.5,0)$ and $\tilde{\bm{t}} = (0,0,0.5)$.
When $\sping$ is reduced to the spin point group $\spinpg$, the orbital part $\mpg$ is centrosymmetric and gray as
                \begin{equation}
                \mpg (\spinpg) = 2/m 1',
                \end{equation}
in contrast to the SO-coupled ($\mpgsoc = 2$). 
Consequently, the spin group symmetry forbids various physical phenomena activated by the time-reversal or space-inversion symmetry breaking. 

Next, we consider a noncoplanar magnet Mn$_3$CuN (\#\,2.5)~\cite{Fruchart1978-df}.
The complex spin structure consists of magnetic moments at Mn sites having two different moduli [Fig.~\ref{Fig_spinsg_examples}(b)].
The magnetic space group is
                \begin{equation}
                \msg = P4/m,
                \end{equation}
from which the magnetic point group is $\mpgsoc = 4/m$.
Thus, owing to the spin order and SOC, the cubic crystalline symmetry (space group No.~221, $Pm\bar{3}m$) is reduced to the tetragonal.
The magnetic symmetry allows for the magnetization along the $[001]$ axis.

Then, let us consider its spin space group symmetry.
With the non-zero propagation vector of the spin configuration, the spin space group includes the spin translation group $\overline{\sping}_\text{st}$ generated by
                \begin{equation}
               \sop{(1,\bm{t})}{2_z},
                \end{equation}
with $\bm{t} = (0.5,0.5,0)$ and by trivial translation operations without any spin-space rotations.
Then, we obtain the coset decomposition of the spin space group as
                \begin{equation}
                \sping  = \bigcup_{i} g_i\, \overline{\sping}_\text{st},
                \label{spin_space_group_Mn3CuN}
                \end{equation}
where the representatives $g_i$ are given by the identity and
                \begin{equation}
                \sop{(-1,\bm{0})}{1},~\sop{(4^+_{[001]},\bm{0})}{4^+_{[001]}},~\sop{(m_{(010)},\bm{0})}{m_{\alpha}},~\sop{(m_{(100)},\bm{0})}{m_{\beta}}.
                \end{equation}
The mirror operation $W =m_{\alpha}$ is depicted in Fig.~\ref{Fig_spinsg_examples}(c).
Importantly, the spin-space-group operations related to $m_{\alpha},m_{\beta}$ are preserved if without the SOC condition of Eq.~\eqref{SOC_constraint} and makes the orbital part $\mpg$ black-white.
The spin point group is obtained as
                \begin{equation}
                \spinpg = \spgop{2_{[001]}}{1} \spgop{4_{[001]}}{4} / \spgop{1}{m}\spgop{m_{\beta}}{m}\spgop{m_{\gamma}}{m}, 
                \end{equation}
with the spin-space mirror operation $W = m_\gamma$ associated with the orbital-space mirror reflection $R = m_{(1\bar{1}0)}$.
Accordingly, we obtain
                \begin{equation}
                \spog (\spinpg) = 4mm,
                \end{equation}
for the spin part, and 
                \begin{equation}
                \mpg (\spinpg) = 4/mm'm',
                \end{equation}
for the orbital part.
The resulting black-white symmetry of $\mpg$ differs from the colorless magnetic point group $\mpgsoc = 4/m$ for the SO-coupled case.

\subsection{Emergent physical properties}

\subsubsection{Magnetization}

Here we consider the spin and orbital magnetization arising from the spin ordering.
Although the uniform spin magnetization trivially appears in the ferromagnetic materials, the orbital counterpart is severely forbidden due to the spin-only group in simple spin structures such as collinear and coplanar configurations (see Sec.~\ref{SecSub_magnetization}).
Then, we focus on the 195 noncoplanar magnets which may possess orbital magnetization.
We show the classification concerning magnetization in Table~\ref{Table_magnetization_magndata}.
The classification also covers the magnetization identified by the magnetic symmetry ($\msg,\mpgsoc$) including the SOC effect, that is, the spin-orbital-entangled magnetization denoted by $\bm{M}_\text{SOC}$.
When either nonzero spin or orbital magnetization exists in a given spin space group, $\bm{M}_\text{SOC}$ is similarly allowed due to the group-subgroup relation of $\msg < \sping$.
Then, we classify the noncoplanar magnets into four classes in terms of magnetization; (M1) nonzero spin and orbital magnetization even without SOC, (M2) nonzero spin but zero orbital magnetization without SOC, (M3) zero spin but nonzero orbital magnetization without SOC, and (M4) zero magnetization without SOC, but nonzero with SOC.
We do not discuss the case of zero magnetization with and without SOC.

The system with orbital-free spin magnetization (Class M2 of Table~\ref{Table_magnetization_magndata}) is trivial since such type of magnetization can be found in typical ferromagnetic materials as well.
On the other hand, the spin-free orbital magnetization (Class M3) indicates a nontrivial spin group symmetry hosting the orbital magnetization not to be concomitant with spin magnetization.
The materials of Class M3 are as follows; DyCrWO$_6$ (\#\,0.316), CuB$_2$O$_4$ (\#\,0.431), Fe$_3$F$_8$(H$_2$O)$_2$ (\#\,2.61), TbCrO$_3$ (\#\,2.62), DyCrO$_3$ (\#\,2.63,~\#\,2.64), and MgCr$_2$O$_4$ (\#\,3.4).
The candidate materials are mostly insulators in contrast to the noncoplanar magnetic metal CoTa$_3$S$_6$ discussed in Sec.~\ref{SecSub_magnetization}.
Thus, they may not be promising candidates offering the geometrical Hall effect, whereas the orbital magnetization should participate in similar phenomena for quasiparticles conductive in electrically-insulating materials such as phonon and magnon~\cite{Onose2010-lo}.
The orbital magnetization also plays an important role in various magneto-optical phenomena such as Faraday rotations~\cite{Feng2020-qv}.
Interestingly, the optical response may be tolerant to extrinsic effects such as skew scattering yielding anomalous Hall effect~\cite{Nagaosa2010-dn} and hence it may be a good test bed for investigating the intrinsic role of the orbital magnetization in emergent responses.

We observe that CrSe (\#\,2.35) shows the magnetization if and only if SOC is taken into account (Class M4).
This is because the cubic symmetry of spin space group $\sping$ is reduced to the trigonal magnetic point group under SOC.

We also notice that the spin and orbital contributions to the equilibrium properties may be distinguished even when both are allowed such as in Class M1 of Table~\ref{Table_magnetization_magndata}.
As an example for Class M1 of Table~\ref{Table_magnetization_magndata}, the spin group symmetry of Mn$_3$O$_4$ (\#\,2.52) leads to the spin and orbital magnetizations given by
                \begin{equation}
                \bm{M}_\text{sp} \parallel [010],~\bm{M}_\text{orb} \parallel [001].
                \label{magnetization_Mn3O4}
                \end{equation}
The two perpendicular magnetizations get entangled with each other under the SOC effect, and the magnetization can be in the $(100)$ plane.
Note that we cannot determine the relative orientation of spin-space axes with respect to the orbital-space coordinate system without SOC.
We, however, determined the spin-space axes of Eq.~\eqref{magnetization_Mn3O4} by referring to the spin configuration observed in experiments, and thus the peculiar relation between spin and orbital magnetization may give an implication; \textit{e.g.}, the anomalous Hall effect $\sigma_{xy}$ may be larger than another transverse component $\sigma_{zx}$, because the former may be related to the orbital magnetization $\bm{M}_\text{orb} \parallel [001]$ while the latter results from SOC.

                \begin{table}[htbp]
                \caption{Classification of noncoplanar magnetic materials in terms of spin and orbital magnetization $\bm{M}_\text{sp},\bm{M}_\text{orb}$. $\bm{M}_\text{SOC}$ denotes the magnetization under the SOC effect. ``Num.'' denotes the number of data of magnetic materials belonging to each class.
                The materials without any magnetization are not shown.
                }
                \label{Table_magnetization_magndata}
                \centering
                \setlength{\tabcolsep}{15pt}
                \begin{tabular}{lCCCc}
                        \toprule
                        &\bm{M}_\text{sp}&\bm{M}_\text{orb}&\bm{M}_\text{SOC}&Num.\\
                        \midrule
                        M1 &\checkmark&\checkmark&\checkmark & 69\\
                        M2 &\checkmark&&\checkmark& 2\\
                        M3 &&\checkmark&\checkmark& 7\\
                        M4 &&&\checkmark & 1\\
                        \midrule
                        \text{Total}&&& & 79 \\
                        \bottomrule
                \end{tabular}
                \end{table}

Finally, we comment on the relation between Class M2 and the spin scalar chirality.
The noncoplanar nature can be quantified by the spin scalar chirality vector $\bm{C}$, which shares the same spin crystallographic symmetry as that of the orbital magnetization and geometrical Hall effect.
The quantity is written by
                \begin{equation}
                C_i = \int d\bm{r} \epsilon_{ijk} \bm{s} (\bm{r}) \cdot \left[ \frac{\partial }{\partial r_j} \bm{s} (\bm{r}) \times \frac{\partial }{\partial r_k} \bm{s} (\bm{r}) \right].
                \end{equation}
The importance of the spin scalar chirality has been explored in various materials such as Kagom\'e lattice~\cite{Ohgushi2000-tf,Shindou2001,Tatara2002-as,Hirschberger2021-gk,Ishizuka2021} and magnetic skyrmion crystals~\cite{Neubauer2009-ya,Fujishiro2021-ri}.

The spin scalar chirality may be defined in a lattice system as 
                \begin{equation}
                \overline{C}_i = \sum_{\triangle} \bm{S}_\alpha \cdot \left(\bm{S}_\beta \times \bm{S}_\gamma  \right)       \left(  \left\{ \alpha,\beta,\gamma \right\} \in \triangle  \right),
                \label{spin_scalar_chirality_lattice}
                \end{equation}
where three spins spanning the triangle ($\triangle$) are in the same plane normal to the $x_i$ axis~\cite{Barros2013-py}.
The triangular unit may be hard to identify in general except for known examples such as layered material~\cite{Takagi2023-ih} and pyrochlore magnet~\cite{Taguchi2001-kj,Hirschberger2021-gk}.
On the other hand, our symmetry analysis allows us to identify the orbital magnetization $\bm{M}_\text{orb}$ as well as the spin scalar chirality $\bm{C}$ in continuum space.
It implies that our symmetry analysis unambiguously identifies the geometrical Hall effect for complex magnetic materials where the discretely defined spin chirality may be hard to identify.

\subsubsection{Magnetic quadrupole moment}

We consider the spin and orbital quadrupole moments $\mq_{ij}^\text{sp}$ and $\mq_{ij}^\text{orb}$.
While the orbital contribution similarly does not show up without a noncoplanar spin structure, the spin contribution is of interest in the low-dimensional spin structures.
Furthermore, we can gain insight into relativistic effects on the quadrupole-mediated physical responses such as the magnetoelectric effect from the comparison between the magnetic quadrupole moments with and without SOC.
Thus, we categorize the magnetic materials by the spin/orbital/SO-coupled ($\mq_{ij}^\text{SOC}$) quadrupole moments and by the spin-structure dimension (Table~\ref{Table_quadrupole_magndata}).

                \begin{table}[htbp]
                \caption{Classification of magnetic quadrupole moments. The spin ($\mq_\text{sp}$), orbital ($\mq_\text{orb}$), and spin-orbital-coupled contributions ($\mq_\text{SOC}$) are classified by the spin-structure dimension ($\mathcal{D}_\text{sp} = 1,2,3$).
                The materials without any quadrupole moments are not shown.
                }
                \label{Table_quadrupole_magndata}
                \centering
                \setlength{\tabcolsep}{15pt}
                \begin{tabular}{lCCCccc}
                 \toprule
                 &\multirow{2}{*}{$\mq_\text{sp}$}&\multirow{2}{*}{$\mq_\text{orb}$}&\multirow{2}{*}{$\mq_\text{SOC}$}& \multicolumn{3}{c}{$\mathcal{D}_\text{sp}$} \\
                 &&&& 1 & 2& 3 \\
                 \midrule
                 Q1 &\checkmark&\checkmark&\checkmark& 0 & 0 & 42\\
                 Q2 &\checkmark&&\checkmark& 142 & 94 & 0\\
                 Q3 &&\checkmark&\checkmark& 0 & 0 & 4\\
                 Q4 &&&\checkmark& 93 & 24 & 1\\
                 \midrule
                \text{Total}&&& & 235 & 212 & 47 \\
                 \bottomrule
                \end{tabular}
                \end{table}

Being consistent with the spin-only-group symmetry for $\spdim = 1$ and $\spdim = 2$, the quadrupole moments originate from only the spin degree of freedom (Class Q2) without SOC for low-dimensional spin structures, while both spin and orbital contributions are admixed (Class Q1) in noncoplanar case ($\spdim = 3$).
Interestingly, only the orbital part is allowed in the SOC-free manner (Class Q3) for U$_3$As$_4$ (\#\,0.169), U$_3$P$_4$ (\#\,0.170), CrSe (\#\,2.35), MgCr$_2$O$_4$ (\#\,3.4).
It is noteworthy that the noncoplanar magnet CrSe possesses the orbital quadrupole moment without any uniform magnetization (see Appendix~\ref{SecApp_classification_noncoplanar}).
The Class Q4 of Table~\ref{Table_quadrupole_magndata} indicates the system whose magnetic quadrupole moment appears if and only if SOC is included.
The magnetoelectric property is inactive without the SOC effect in some collinear antiferromagnets because of strong constraints from the spin-only group.

The magnetoelectric effect related to magnetic quadrupole moment has been intensively studied mainly with collinear magnets~\cite{Fiebig2005-hj} and with incommensurate magnetic systems~\cite{Kimura2003-ma,Tokura2014-ix}.
It, however, has been rarely explored for commensurate but noncollinear magnetic materials such as what manifests the orbital magnetoelectric effect~\cite{Pimenov2006-zu,Delaney2009-od,Kimura2018-vr}.
Thus, the present classification may lead us to a deep comprehension of the relation between the spin-structure dimension and the magnetoelectric effect.

\subsubsection{Rotators for the spin-polarized current responses}

Finally, let us consider the T-even and T-odd spin-current rotators $R_{ij}^\text{e,o}$.
These quantities are comprised of the spin degree of freedom in $j$-th component of $R_{ij}$ and vanishes in the paramagnetic state without SOC.
Under the SOC effect, the T-even rotator $R_{ij}^\text{e}$ is allowed in every system as in Eq.~\eqref{SOC_and_spinHall}, while the T-odd rotator $R_{ij}^\text{o}$ requires the time-reversal symmetry breaking~\footnote{To be more precise, the T-odd rotator requires the axial symmetry and violation of the combined symmetry of the space-inversion and time-reversal symmetry in addition to the time-reversal-symmetry breaking.}.
Then, we identify candidate materials possessing the T-even rotator without SOC ($R_{ij}^\text{e}$), T-odd rotator without SOC ($R_{ij}^\text{o}$), and T-odd rotator under the SOC effect ($R_{ij}^\text{oS}$) for each spin-structure dimension (Table~\ref{Table_rotators_magndata}).

                \begin{table}[htbp]
                \caption{Classification of the spin-current rotators. Magnetic material data with the spin-structure dimension $\mathcal{D}_\text{sp} = 1,2,3$ are classified in terms of the T-even ($R_\text{e}$), T-odd ($R_\text{orb}$), and SO-coupled T-odd rotators ($R_\text{oS}$).
                The materials without any rotators are not shown.
                }
                \label{Table_rotators_magndata}
                \centering
                \setlength{\tabcolsep}{15pt}
                \begin{tabular}{lCCCccc}
                \toprule
                & \multirow{2}{*}{$R_\text{e}$}&\multirow{2}{*}{$R_\text{o}$}&\multirow{2}{*}{$R_\text{oS}$}& \multicolumn{3}{c}{$\mathcal{D}_\text{sp}$} \\
                &&&& 1 & 2& 3 \\
                \midrule
                R1 &\checkmark&\checkmark&\checkmark & 0 & 115 & 100\\
                R2 &&\checkmark&\checkmark& 130 & 19 & 0\\
                R3 &&&\checkmark     & 123 & 42 & 5\\
                R4 &\checkmark&& & 0 & 164 & 79\\
                R5 &\checkmark&&\checkmark& 0 & 5 & 4\\
                \midrule
                \text{Total}&&& & 253 & 345 & 188 \\
                \bottomrule
                \end{tabular}
                \end{table}

For the collinear case ($\spdim = 1$), the T-even rotator vanishes due to the spin-only group symmetry ($R_{ij}^\text{e} =0$), but the T-odd contribution can be finite (Class R2 and R3).
As a result, the spin Hall effect of collinear magnets is generically attributed to the magnetic origin and hence is unique to magnetic metals.
It is noteworthy that the spin current induced by the T-odd spin Hall effect is not accompanied by the charge current in the collinear and coplanar magnets due to the absence of the anomalous Hall effect without SOC.
This situation is distinct from the Hall effect of SO-coupled systems where the magnetic spin Hall current is admixed with the charge Hall current~\cite{Naka2021-ow}.

For coplanar and noncoplanar magnets ($\spdim=2,3$), both the electric and T-odd rotators are not forbidden in general.
In particular, the candidates for Class R5 may show a sizable T-even spin Hall effect without the help of SOC as demonstrated in the first-principles study such as that for Mn$_3$Sn~\cite{Zhang2018-yj}.
On the other hand, to be different from the collinear case, the T-odd rotator rarely appears without being admixed with the T-even rotator in noncoplanar magnets (Class R2 and R3), because a complex spin structure providing the T-odd rotator secondarily induces the time-reversal-symmetric spin-charge anisotropy as well.

We note that the symmetry analysis refers to the spin structures reported in experiments, that is, the spin configurations including the SOC effect such as small canting by the Dzyaloshinskii-Moriya interaction.
To obtain a proper insight into the SOC-free responses, we have to remove the SOC corrections by performing the comparative and first-principles study with and without the SOC effect~\cite{Huebsch2021-as,Yanagi2023-ci}.
The noncollinear spin configuration does not necessarily require the SOC effect because the collinear configuration is not favorable in some cases such as in frustrated systems~\cite{Yoshimori1978-wn,Hatanaka2023-dc}.
More exploration of the SOC-free magnet and its emergent responses is a future work to be addressed.

\section{Discussion and Summary}
\label{Sec_discussions}

We mainly focused on linear responses such as anomalous Hall, magnetoelectric, and spin Hall responses.
The powerful features, such as T-even/T-odd decomposition and classification in terms of spin and orbital degrees of freedom, work in analyzing nonlinear responses as well.
We exemplify it by nonreciprocal DC current induction in \cota{}.
The response reads as
                \begin{equation}
                J_i(\omega=0) = \sigma_{i;jk} E_j (\omega_0) E_k (-\omega_0).
                \label{nonreciprocal_conductivity_formula}
                \end{equation}
It is called photocurrent response for $\omega_0 \neq 0$~\cite{Ma2021-xq} and nonreciprocal conductivity for $\omega_0 = 0$~\cite{Tokura2018-ht}.
Performing the T-even and T-odd decompositions, we obtain the allowed components
                \begin{equation}
                \sigma_{z;xy} = -\sigma_{z;yx},~\sigma_{y;zx} = -\sigma_{y;xz},~\sigma_{x;yz} = -\sigma_{x;zy},
                \end{equation}
for the T-even contribution and 
                \begin{equation}
                \sigma_{z;xx} = \sigma_{z;xx},~\sigma_{x;zx} = \sigma_{y;zy},~\sigma_{x;xz} = \sigma_{y;yz},~\sigma_{z;zz},
                \end{equation}
for the T-odd.
Although the T-even components are due to the noncentrosymmetric crystal structure of CoTa$_3$S$_6$, the T-odd component is correlated with emergence of the orbital magnetic toroidal moment [see Sec.~\ref{SecSub_geometric_magnetic_multipoles} and Fig.~\ref{Fig_magnetoelectric}(b)]~\cite{Rikken2005-sp,Zelezny2021-wm,Yatsushiro2022-ui,Liu2022-se}.
This implies the giant nonlinear response driven by the nonrelativistic spin-charge coupling~\cite{Hayami2022-wt}.
Supporting this argument, it has been identified that the giant exchange splitting leads to the sizable nonreciprocal conductivity in Mn-based antiferromagnetic metals~\cite{Watanabe2020-oe}.

Furthermore, we can separate the spin and orbital contributions for the magnetization-related nonlinear responses.
By replacing the dc electric current in Eq.~\eqref{nonreciprocal_conductivity_formula} with the spin-/orbital-polarized DC $J_i^{s_a}$/$J_i^{L_a}$, the symmetry analysis can be applied to the nonreciprocal spin/orbital current induction $\sigma_{i;jk}^{a,\text{sp}}$/$\sigma_{i;jk}^{a,\text{orb}}$.
Owing to the high symmetric spin space, the spin part vanishes ($\sigma_{i;jk}^{a,\text{sp}} = 0$) but the orbital contribution exists ($\sigma_{i;jk}^{a,\text{orb}}\neq 0$) in the case of \cota{}.
The situation is different from that considered in the previous study on the spin contribution~\cite{Liu2023-qg}.

Our work provides a systematic tool for investigating SOC-free responses, whereas it does not address quantitative aspects of responses due to the limitation of symmetry analysis.
Previous studies reported that the SOC-free physical response can be sizable such as spin-polarized current induction~\cite{Zhang2018-yj,Gonzalez-Hernandez2021-pb} and piezomagnetic effect~\cite{Ma2021-ji}, but criteria for identifying giant spin-driven phenomena remain elusive.
These problems are expected to be addressed in future works as guided by our symmetry analysis.
For instance, the classification of the magnetic quadrupole moments (Table~\ref{Table_quadrupole_magndata}) may motivate us to revisit the magnetoelectric effect.
Historically, the effect has been mainly explored with simple and collinear magnets such as Cr$_2$O$_3$.
For collinear magnets without SOC, the one-dimensional spin-only group allows for only the longitudinal magnetoelectric effect where induced magnetization is collinear to the spins as in Eq.~\eqref{Cr2O3_magnetoelectricity}.
Given that small longitudinal spin fluctuations suppress the magnetoelectric effect at the low temperature~\cite{Hornreich1967-bo}, the SOC-free magnetoelectric effect is typically small for the collinear magnets without thermal fluctuations.
On the other hand, coplanar and noncoplanar magnetic materials may host significant spin magnetoelectric responses due to remaining spin fluctuations.
Although the importance of the noncollinear property has been highlighted by prior theoretical studies~\cite{Delaney2009-od}, candidate materials are not fully explored.
The developed spin-group symmetry analysis incorporated into the computational design of magnetic materials~\cite{Huebsch2021-as} may facilitate further investigations into complex spin structures enhancing magnetoelectric responses.

Previous theoretical studies investigated the spin-momentum coupling originating from the spin order without the SOC effect~\cite{Hayami2020-gk,Yuan2020-kx,Yuan2021-el,Turek2022-aa,Smejkal2022-ir,Yang2021-vn}.
The purpose of this paper is to clarify the macroscopic physical properties, and the spin-space-group analysis of the spin-splitting structure is out of the present scope.
The systematic classification, however, can be similarly obtained by combining the spin-space-group symmetry with the classification of spin momentum locking in the light of multipolar degrees of freedom~\cite{Watanabe2017-qk,Watanabe2018-cu,Hayami2018-bh}.
By properly taking the characteristics of the spin space group, we can get accurate criteria for spontaneous spin-momentum splitting.
Interestingly, various emergent physical responses can occur even without the nonrelativistic spin-momentum locking such as exemplified by the geometrical Hall effect of \cota{} because of the spin-translation degree of freedom.

To summarize, we established the spin-crystallographic-group symmetry analysis applicable to complex spin structures such as that with a nontrivial spin translation group.
Our symmetry analysis is powerful enough to cultivate further understandings of spin-order-induced emergent responses and relativistic corrections to them in various magnetic materials.
In stark contrast to the widely adopted magnetic space group, the spin space group takes into account the spin-structure dimension and conveniently allows us to identify the geometric contributions to physical responses such as the geometrical Hall effect and orbital magnetoelectric effect.
The spin-space-group symmetry and characteristic physical responses are automatically identified by our computational methods developed in Ref.~\cite{Shinohara2024-ld} and in this work.
Performing the computational classification of dozens of magnetic materials, we systematically identified intriguing systems such as what hosts purely-orbital magnetization, purely-orbital magnetic quadrupole polarization, and the spin-order-induced T-even/T-odd spin Hall effect.
The developed symmetry analysis will deepen our understanding of spin-orbit-free phenomena by combining first-principles material design.

The program~\footnote{\url{https://github.com/Hi-Wat/scg-symmetry-search.git}} of searching the symmetry-adapted tensors with a given spin space group is based on \textsc{spglib}~\cite{togo2018tspglib,Shinohara2023-sy} and \textsc{spinspg}~\cite{Shinohara2024-ld}, and it is distributed under the BSD 3-clause license.
The program allows for the T-even/T-odd decomposition and supports various physical properties such as equilibrium property, linear response, and nonlinear responses.

\noindent\textit{Note added---}

Recently, we noticed Refs.~\cite{Jiang2023-ad,Xiao2023-lx,Ren2023-av} relevant to our work.
These works worked on the identification and classification of spin space groups.

\section*{acknowledgement}
The authors are grateful Rikuto Oiwa and Susumu Minami for fruitful comments and discussions.
This work is supported by JSPS-KAKENHI (No.~JP23K13058, JP22H00290, JP21H04437, JP21H04990, JP21J00453, JP21J10712, and JP19H05825), JST-CREST Program (No.~JPMJCR18T3, JPMJCR23O4), and JST-PRESTO (No.~JPMJPR20L7).
The crystal structure and its spin configuration are visualized by a useful software \textsc{vesta}~\cite{Momma2011-jl}.

\appendix

\section{Notes on group theory}
\label{SecApp_note_group_theory}

\subsection{General properties of group theory}

The terminology used in the paper is briefly mentioned to make the paper self-contained.
Let $g = \sop{h}{W},g' = \sop{h'}{W'} $ be operations of the spin space group $\sping$, the multiplication law is defined as
                \begin{equation}
                g \cdot g' = \sop{h\cdot h'}{W\cdot W'},
                \end{equation}
where the multiplications of the spin- and orbital-space operations are similarly defined in that for the O(3) group and for the space group, respectively.
Accordingly, the spin space group satisfies the group axioms, that is, associativity ($g\cdot g' \in \sping$), the existence of identity (id.) operation (for $\text{id.} \in \sping$, $g \cdot \text{id.}  = \text{id.}  \cdot g = g$), and the existence of the inverse operation (for $g\in \sping$, there uniquely exists the operation $g^{-1} \in \sping$ such that $g\cdot g^{-1} = g^{-1} \cdot g = \text{id.}$).
We denote $|G|$ as the order of a group $G$ (the number of operations in $G$).

Let $H$ be a group whose all the operations are in $G$, and the group-subgroup relation holds as $G > H$.
The group-subgroup relation indicates that the group $G$ can be decomposed by its subgroup $H$ as               
                \begin{align}
                G 
                &= g_1 H \cup g_2 H \cup \cdots \cup g_n H,\\ 
                &= \bigcup_i g_i H.
                \label{right_decomposition}
                \end{align}
Representatives $g_i \in G$ are chosen to satisfy the relation
                \begin{equation}
                g_i H \cap g_j H = \phi ~(i \neq j),
                \end{equation}
indicating that every intersection is empty ($\phi$).
The decomposition is similarly performed from the right-hand side as
                \begin{equation}
                G = \bigcup_i H \, \overline{g}_i.
                \label{left_decomposition}
                \end{equation}

Let us consider the subgroup $H$ commuting with every operation of its supergroup $G$ as 
                \begin{equation}
                g \in G,~ g  H = H  g,
                \end{equation}
then $H$ is a \textit{normal subgroup} of $G$ denoted as
                \begin{equation}
                H  \triangleleft G.
                \end{equation}
A trivial normal subgroup is $G$ itself ($G \triangleleft G$).
If $H \triangleleft G$ holds, two-fold coset decompositions in Eqs.~\eqref{right_decomposition} and \eqref{left_decomposition} are equivalent.
In that case, the cosets $\{g_i H \}$ form the factor group $G/H$ whose multiplication law is
                \begin{equation}
                g_i H \cdot g_j H = (g_i g_j) H.
                \end{equation}
 
While the representatives of the factor group $\{g_i\}$ themselves do not form a group in general, there may exist the group $K = \{g_i\}$ satisfying the group axiom such as $g_i \cdot g_j \in K$.
Accordingly, $G$ is recast as the \textit{internal semidirect product} of $H$ and K, 
                \begin{equation}
                G = H \rtimes K,
                \end{equation}
where $H,K < G$ , $H \triangleleft \, G$, and there exists the trivial intersection between $H$ and $K$ as $K \cap H = \{ \text{id.} \}$.
When $G =H \rtimes K  $ holds, the operations in $G$ can be written as
                \begin{equation}
                G = \left\{ h \cdot k \, | h \in H,~ k \in K\right\}.
                \label{group_element_internal_semidirect}
                \end{equation}
If we further demand the relation $K \triangleleft \, G$, the group $G$ is given by the \textit{(internal) direct product}
                \begin{equation}
                G = H \times K,
                \end{equation}
by which the operations $h \in H $ and $k \in K$ commute with each other in Eq.~\eqref{group_element_internal_semidirect}.

\subsection{Group structure of spin space group}

The spin space group $\sping$ has a hierarchical structure given as follows~\cite{Litvin1973-ii}.
The entire group can be recast as the coset decomposition with the spin translation group ($\sping_\text{st} \triangleleft \, \sping$)  
                \begin{align}
                \sping   = \bigcup_i g_i \, \sping_\text{st}.
                \label{coset_decomp_sping_spinst}
                \end{align}
The spin translation group is given by
                \begin{equation}
                        \sping_\text{st} = \left\{  \sop{(1,\bm{t})}{W}  \in \sping  \right\},
                \end{equation}
containing spin-only operations and combinations of the spin rotation and translation such as $\sop{(1,\bm{0})}{W}, \sop{(1,\bm{t})}{W}$.
The spin translation group is further divided as~\cite{Litvin1973-ii}
                \begin{equation}
                \sping_\text{st} = \sping_\text{so} \times  \overline{\sping}_\text{st}.
                \label{spinst_to_spinonly_nontrivial_spinst}
                \end{equation}
The spin only group $\sping_\text{so}$ is comprised of only the spin rotation operation
                \begin{equation}
                \sping_\text{so} = \left\{   \sop{(1,\bm{0})}{W}  \in \sping  \right\},
                \end{equation}
in which the point group operations $\{ W \}$ form the spin-only group $\spinpg_\text{so}$.
A group $\overline{\sping}_\text{st}$ in Eq.~\eqref{spinst_to_spinonly_nontrivial_spinst} denotes the nontrivial spin translation group whose spin rotation should be coupled to the translation. 

We can perform the decomposition of the spin space group in a different manner from that in Eq.~\eqref{coset_decomp_sping_spinst} as
                \begin{equation}
                \sping = \sping_\text{so} \times \overline{\sping},
                \label{sping_to_spinonly_nontrivialspinst} 
                \end{equation}
by which the nontrivial spin space group $\overline{\sping}$ is defined.
The nontrivial spin space group contains the nontrivial spin translation group as its normal subgroup ($\overline{\sping}_\text{st} \triangleleft \, \overline{\sping}$).
Thus, the coset decomposition is obtained as
                \begin{equation}
                        \overline{\sping} = \bigcup_i g_i\,  \overline{\sping}_\text{st}.
                        \label{coset_decomp_nontrivialsping_spinst}
                \end{equation}
In the main text, we mainly discuss the spin space group by using the decomposition of Eq.~\eqref{sping_to_spinonly_nontrivialspinst}.

The nontrivial spin space group can be given by the internal semidirect product in some cases; \textit{e.g.}, \cota{} of Eq.~\eqref{spin_space_group_CoTa3S6}.
It is, however, not always the case.
For instance, $\overline{\sping}$ in Eq.~\eqref{spin_space_group_Mn3CuN} cannot be given that way.
See also Ref.~\cite{Shinohara2024-ld}.
The representatives of the factor group $\overline{\sping}/\overline{\sping}_\text{st}$ are $g = \sop{(R,\bm{t})}{W}$ such that $R \neq 1$ otherwise $g$ is the identity.
The factor group $\left\{ g_i\,  \overline{\sping}_\text{st} \right\}$ is therefore isomorphic to a nontrivial spin point group $\spinpg_\mathcal{H}$~\cite{Litvin1977,Liu2022-dh} defined by
                \begin{equation}
                \spinpg_\mathcal{H} = \{\sop{R}{W}| \sop{R}{W} \neq \sop{1}{W} \text{ for } W \neq 1 \}.
                \end{equation}

\subsection{Conventional classification of magnetic symmetry}

The magnetic space and point groups account for the magnetic symmetry of nonmagnetic and magnetic systems~\cite{bradley2010mathematical}.
In the following, we consider widely-used magnetic symmetry respecting the SOC constraint.
The group consists of symmetry operations with and without the time-reversal operation which are respectively unitary and anti-unitary, while it may contain only unitary operations in some cases.
In particular, when there exists an anti-unitary operation in group $G$, $G$ has a normal subgroup $H$ consisting of only unitary operations whose order is half that of $G$ ($|H| = |G|/2$).
Then, we obtain the coset decomposition
                \begin{equation}
                G = H \cup a H,
                \label{black_white_decomposition}
                \end{equation}
where $a \not\in H$ is an anti-unitary operation including the time-reversal operation $\theta \equiv 1'$.
Otherwise, the group is formed by only the unitary operations.
In terms of anti-unitary symmetry, magnetic symmetry is classified as follows.

\subsection*{Magnetic point group}

The magnetic point group is comprised of orbital-space unitary ($R$) and anti-unitary ($R' = \theta R$) operations where $R$ belongs to the O(3) group.
The magnetic point groups $\mpgsoc$ are classified into three types;
\begin{description}
        \item[Colorless group] no element including the time-reversal operation in $\mpgsoc$,
        \item[Gray group] the time-reversal symmetry trivially holds as $1' \in \mpgsoc$,
        \item[Black-White group] otherwise, \textit{i.e.,} $a = R'$ with $R\neq 1$ in Eq.~\eqref{black_white_decomposition}.
\end{description}

\subsection*{Magnetic space group}

The magnetic space group $\msg$ is formed by the point-group operation $R$ ($R'$) without (with) the time-reversal operation and by the translation operation $\bm{t}$.
These two operations are frequently summarized to the Seitz notation such as $(R,\bm{t}),~(R',\tilde{\bm{t}})$.

Similarly to the magnetic point group, the magnetic space groups are classified into four types with respect to the anti-unitary symmetry~\cite{Litvin1977}.
\begin{description}
        \item[Type I] no element including the time-reversal operation in $\msg$,
        \item[Type II] the time-reversal symmetry trivially holds as $(1',\bm{0}) \in \msg$,
        \item[Type III] the time-reversal operation is combined with the point-group operation as $a = (R',\bm{t})$ with $R\neq 1$ in Eq.~\eqref{black_white_decomposition},
        \item[Type IV] the time-reversal operation is combined with the translational operation as $a = (1',\tilde{\bm{t}})$.
\end{description}
Note that the translation $\tilde{\bm{t}}$ belongs to the translational group $\mathcal{T}_0 = \left\{ (1,\bm{t})\right\}$ for the paramagnetic phase but is not included in that for the magnetic state.

Magnetic materials do not show the trivial time-reversal symmetry related to $g = (1',\bm{0})$.
Thus, we obtain the one-to-one correspondence between the types of the magnetic point group and magnetic space group as $\left( \mpgsoc, \msg \right)$ = (Colorless, I), (Gray, IV), (Black-White, III).

\section{Canonical correlation for linear response theory and its symmetry constraint}
\label{SecApp_canonical_correlation}

Let us introduce the linear response function of $X_i = \chi_{ij}^{XY} F_j^{(Y)}$ defined in Eq.~\eqref{linear_response_formula} in the frequency domain as
                \begin{equation}
                \chi_{ij}^{XY} (\omega) = \int_0^{\infty} dt e^{i\omega t -\eta t} \chi_{ij}^{XY} (t),
                \end{equation}
with the infinitesimal positive parameter $\eta =+0$  building the causality into the response.
The response function in the time domain can be written as the form of canonical correlation as~\cite{kubo2012statistical}
                \begin{equation}
                \chi_{ij}^{XY} (t) \equiv \Gamma_{ij}^{X\dot{Y}}= \int_0^{1/T} d\tau \text{Tr}\left[ \rho_\text{eq} \dot{Y}_j (-i\tau) X_i (t)  \right].
                \end{equation}
The operators are in the Heisenberg representation, $X (t) = e^{iH t} X e^{-iHt}$ with the unperturbed Hamiltonian $H$.
We also introduced the temperature $T$ and the density operator for the (unperturbed) equilibrium state $\rho_\text{eq} = e^{-H/T} / \text{Tr}\left[ e^{-H/T} \right]$.
Following Ref.~\cite{Seemann2015-ns}, the transformation property of the canonical correlation function (in the frequency domain) is 
                \begin{equation}
                        \Gamma_{ij}^{X\dot{Y}} (\omega) = \Gamma_{kl}^{X\dot{Y}} (\omega ) D_{ki}^{(\bm{X})} (g) D_{lj}^{(\bm{Y})} (g),
                \end{equation}
for a preserving unitary operation $g$ and 
                \begin{equation}
                        \Gamma_{ij}^{X\dot{Y}} (\omega) = - \Gamma_{kl}^{\dot{Y}X} (\omega ) \left[ D_{ki}^{(\bm{X})} (g) \right]^\ast  \left[ D_{lj}^{(\bm{Y})} (g) \right]^\ast,
                \end{equation}
for an anti-unitary operation.
Let us consider the electric conductivity by adopting the electric current $\bm{X} = \bm{J}$ and the electric polarization $\bm{Y} = \bm{P}$.
When the orbital time-reversal symmetry $g = \sop{1}{W}$ (det\,$W=-1$) is preserved in a given magnetic group such as spin point group $\spinpg$ and SO-coupled magnetic point group $\mpgsoc$, we can relate different components of the electric conductivity $\sigma_{ij} = \chi_{ij}^{JP}$ with each other as 
                \begin{equation}
                \sigma_{ij} =\chi_{ij}^{JP} = \kappa_{ij}^{JJ} = - \kappa_{kl}^{JJ} \left( -\mathds{1} \right)_{kj} \left( \mathds{1} \right)_{li} = \kappa_{ji}^{JJ} = \chi_{ji}^{JP} = \sigma_{ji},
                \end{equation}
by using $\dot{\bm{P}} = \bm{J}$.
This means the Onsager reciprocity.

\section{Classification of noncoplanar magnets}
\label{SecApp_classification_noncoplanar}
We summarize the classification of noncoplanar magnets shown in Sec.~\ref{Sec_magndata_study}.
For details of the definitions of each quantity, please refer to the corresponding tables (Tables~\ref{Table_magnetization_magndata},~\ref{Table_quadrupole_magndata},~\ref{Table_rotators_magndata}).

\begin{longtable*}{lccccccccc}
\caption{Classification of noncoplanar magnets in terms of magnetization ( $\bm{M}_\text{sp}$, $\bm{M}_\text{orb}$, $\bm{M}_\text{SOC}$), magnetic quadrupole moment ( $\mathrm{Q}_\text{sp}$, $\mathrm{Q}_\text{orb}$, $\mathrm{Q}_\text{SOC}$), and rotators ( $R_\text{e}$, $R_\text{o}$, $R_\text{oS}$). The names of each material refer to \textsc{magndata}.}
\label{Table_noncoplanar_classification_summary}\\

\toprule
\# & $\bm{M}_\text{sp}$ & $\bm{M}_\text{orb}$ & $\bm{M}_\text{SOC}$ & $\mathrm{Q}_\text{sp}$ & $\mathrm{Q}_\text{orb}$ & $\mathrm{Q}_\text{SOC}$ & $R_\text{e}$ & $R_\text{o}$ & $R_\text{oS}$ \\ 
\midrule
\endfirsthead

\multicolumn{10}{c}{\tablename\ \thetable\ (\textit{cont.})} \\
\# & $\bm{M}_\text{sp}$ & $\bm{M}_\text{orb}$ & $\bm{M}_\text{SOC}$ & $\mathrm{Q}_\text{sp}$ & $\mathrm{Q}_\text{orb}$ & $\mathrm{Q}_\text{SOC}$ & $R_\text{e}$ & $R_\text{o}$ & $R_\text{oS}$ \\ 
\midrule
\endhead

\endfoot

\endlastfoot

\verb|0.102_Mn2GeO4.mcif| & &  &  &  &  &  & $\checkmark$ & $\checkmark$ & $\checkmark$ \\ 
\verb|0.103_Mn2GeO4.mcif| &$\checkmark$ & $\checkmark$ & $\checkmark$ &  &  &  & $\checkmark$ & $\checkmark$ & $\checkmark$ \\ 
\verb|0.106_DyVO3.mcif| &$\checkmark$ & $\checkmark$ & $\checkmark$ &  &  &  & $\checkmark$ & $\checkmark$ & $\checkmark$ \\ 
\verb|0.127_Dy3Al5O12.mcif| & &  &  &  &  &  & $\checkmark$ &  &  \\ 
\verb|0.135_Ni3B7O13Br.mcif| &$\checkmark$ & $\checkmark$ & $\checkmark$ & $\checkmark$ & $\checkmark$ & $\checkmark$ & $\checkmark$ & $\checkmark$ & $\checkmark$ \\ 
\verb|0.136_Co3B7O13Br.mcif| &$\checkmark$ & $\checkmark$ & $\checkmark$ & $\checkmark$ & $\checkmark$ & $\checkmark$ & $\checkmark$ & $\checkmark$ & $\checkmark$ \\ 
\verb|0.141_Tb5Ge4.mcif| & &  &  & $\checkmark$ & $\checkmark$ & $\checkmark$ & $\checkmark$ &  &  \\ 
\verb|0.145_Co3TeO6.mcif| & &  &  & $\checkmark$ & $\checkmark$ & $\checkmark$ & $\checkmark$ &  &  \\ 
\verb|0.150_NiS2.mcif| & &  &  &  &  &  & $\checkmark$ & $\checkmark$ & $\checkmark$ \\ 
\verb|0.151_Tm2Mn2O7.mcif| &$\checkmark$ & $\checkmark$ & $\checkmark$ &  &  &  & $\checkmark$ & $\checkmark$ & $\checkmark$ \\ 
\verb|0.157_Yb2Sn2O7.mcif| &$\checkmark$ & $\checkmark$ & $\checkmark$ &  &  &  & $\checkmark$ & $\checkmark$ & $\checkmark$ \\ 
\verb|0.158_Yb2Ti2O7.mcif| &$\checkmark$ & $\checkmark$ & $\checkmark$ &  &  &  & $\checkmark$ & $\checkmark$ & $\checkmark$ \\ 
\verb|0.167_Nd3Sb3Mg2O14.mcif| &$\checkmark$ & $\checkmark$ & $\checkmark$ &  &  &  & $\checkmark$ & $\checkmark$ & $\checkmark$ \\ 
\verb|0.168_NH4Fe2F6.mcif| & &  &  &  &  &  & $\checkmark$ & $\checkmark$ & $\checkmark$ \\ 
\verb|0.169_U3As4.mcif| &$\checkmark$ &  & $\checkmark$ &  & $\checkmark$ & $\checkmark$ &  &  & $\checkmark$ \\ 
\verb|0.170_U3P4.mcif| &$\checkmark$ &  & $\checkmark$ &  & $\checkmark$ & $\checkmark$ &  &  & $\checkmark$ \\ 
\verb|0.184_Nd5Si4.mcif| &$\checkmark$ & $\checkmark$ & $\checkmark$ & $\checkmark$ & $\checkmark$ & $\checkmark$ & $\checkmark$ & $\checkmark$ & $\checkmark$ \\ 
\verb|0.185_Nd5Ge4.mcif| &$\checkmark$ & $\checkmark$ & $\checkmark$ &  &  &  & $\checkmark$ & $\checkmark$ & $\checkmark$ \\ 
\verb|0.203_Mn3Ge.mcif| &$\checkmark$ & $\checkmark$ & $\checkmark$ &  &  &  & $\checkmark$ &  & $\checkmark$ \\ 
\verb|0.204_Ca2MnReO6.mcif| &$\checkmark$ & $\checkmark$ & $\checkmark$ &  &  &  & $\checkmark$ & $\checkmark$ & $\checkmark$ \\ 
\verb|0.20_MnTe2.mcif| & &  &  &  &  &  & $\checkmark$ & $\checkmark$ & $\checkmark$ \\ 
\verb|0.218_Co2SiO4.mcif| & &  &  &  &  &  & $\checkmark$ & $\checkmark$ & $\checkmark$ \\ 
\verb|0.219_Co2SiO4.mcif| & &  &  &  &  &  & $\checkmark$ & $\checkmark$ & $\checkmark$ \\ 
\verb|0.220_Mn2SiO4.mcif| &$\checkmark$ & $\checkmark$ & $\checkmark$ &  &  &  & $\checkmark$ & $\checkmark$ & $\checkmark$ \\ 
\verb|0.221_Fe2SiO4.mcif| & &  &  &  &  &  & $\checkmark$ & $\checkmark$ & $\checkmark$ \\ 
\verb|0.236_CaFe4Al8.mcif| & &  &  &  &  &  & $\checkmark$ & $\checkmark$ & $\checkmark$ \\ 
\verb|0.240_Er2Cu2O5.mcif| & &  &  & $\checkmark$ & $\checkmark$ & $\checkmark$ & $\checkmark$ & $\checkmark$ & $\checkmark$ \\ 
\verb|0.250_(NH2(CH3)2)(FeCo(HCOO)6).mcif| &$\checkmark$ & $\checkmark$ & $\checkmark$ &  &  &  & $\checkmark$ & $\checkmark$ & $\checkmark$ \\ 
\verb|0.251_(NH2(CH3)2)(FeMn(HCOO)6).mcif| &$\checkmark$ & $\checkmark$ & $\checkmark$ &  &  &  & $\checkmark$ & $\checkmark$ & $\checkmark$ \\ 
\verb|0.268_Tb2MnNiO6.mcif| &$\checkmark$ & $\checkmark$ & $\checkmark$ & $\checkmark$ & $\checkmark$ & $\checkmark$ & $\checkmark$ & $\checkmark$ & $\checkmark$ \\ 
\verb|0.269_Tb2MnNiO6.mcif| &$\checkmark$ & $\checkmark$ & $\checkmark$ &  &  &  & $\checkmark$ & $\checkmark$ & $\checkmark$ \\ 
\verb|0.281_Co2V2O7.mcif| & &  &  & $\checkmark$ & $\checkmark$ & $\checkmark$ & $\checkmark$ &  &  \\ 
\verb|0.292_NiTe2O5.mcif| & &  &  &  &  &  & $\checkmark$ & $\checkmark$ & $\checkmark$ \\ 
\verb|0.294_Cu4(OD)6FBr.mcif| &$\checkmark$ & $\checkmark$ & $\checkmark$ &  &  &  & $\checkmark$ & $\checkmark$ & $\checkmark$ \\ 
\verb|0.29_Er2Ti2O7.mcif| & &  &  &  &  &  & $\checkmark$ & $\checkmark$ & $\checkmark$ \\ 
\verb|0.2_Cd2Os2O7.mcif| & &  &  &  &  &  & $\checkmark$ &  &  \\ 
\verb|0.311_CoGeO3.mcif| & &  &  & $\checkmark$ & $\checkmark$ & $\checkmark$ & $\checkmark$ &  &  \\ 
\verb|0.316_DyCrWO6.mcif| & & $\checkmark$ & $\checkmark$ & $\checkmark$ & $\checkmark$ & $\checkmark$ & $\checkmark$ & $\checkmark$ & $\checkmark$ \\ 
\verb|0.318_Tm2CoMnO6.mcif| &$\checkmark$ & $\checkmark$ & $\checkmark$ &  &  &  & $\checkmark$ & $\checkmark$ & $\checkmark$ \\ 
\verb|0.326_Nd2Sn2O7.mcif| & &  &  &  &  &  & $\checkmark$ &  &  \\ 
\verb|0.339_Nd2Hf2O7.mcif| & &  &  &  &  &  & $\checkmark$ &  &  \\ 
\verb|0.33_HoMnO3.mcif| & &  &  & $\checkmark$ & $\checkmark$ & $\checkmark$ & $\checkmark$ & $\checkmark$ & $\checkmark$ \\ 
\verb|0.340_Nd2Zr2O7.mcif| & &  &  &  &  &  & $\checkmark$ &  &  \\ 
\verb|0.342_Tb3Ge5.mcif| & &  &  & $\checkmark$ & $\checkmark$ & $\checkmark$ & $\checkmark$ & $\checkmark$ & $\checkmark$ \\ 
\verb|0.347_Er2ReC2.mcif| & &  &  & $\checkmark$ & $\checkmark$ & $\checkmark$ & $\checkmark$ &  &  \\ 
\verb|0.349_Nd2NiO4.mcif| &$\checkmark$ & $\checkmark$ & $\checkmark$ &  &  &  & $\checkmark$ & $\checkmark$ & $\checkmark$ \\ 
\verb|0.352_TbFeO3.mcif| &$\checkmark$ & $\checkmark$ & $\checkmark$ &  &  &  & $\checkmark$ & $\checkmark$ & $\checkmark$ \\ 
\verb|0.357_CaFe5O7.mcif| &$\checkmark$ & $\checkmark$ & $\checkmark$ &  &  &  & $\checkmark$ & $\checkmark$ & $\checkmark$ \\ 
\verb|0.368_(CH3NH3)(Co(COOH)3.mcif| &$\checkmark$ & $\checkmark$ & $\checkmark$ &  &  &  & $\checkmark$ & $\checkmark$ & $\checkmark$ \\ 
\verb|0.369_(CH3NH3)(Co(COOH)3.mcif| &$\checkmark$ & $\checkmark$ & $\checkmark$ &  &  &  & $\checkmark$ & $\checkmark$ & $\checkmark$ \\ 
\verb|0.388_Co3Al2Si3O12.mcif| & &  &  & $\checkmark$ & $\checkmark$ & $\checkmark$ & $\checkmark$ &  &  \\ 
\verb|0.394_Cu2CdB2O6.mcif| & &  &  & $\checkmark$ & $\checkmark$ & $\checkmark$ & $\checkmark$ &  &  \\ 
\verb|0.39_Nd2NaRuO6.mcif| &$\checkmark$ & $\checkmark$ & $\checkmark$ &  &  &  & $\checkmark$ & $\checkmark$ & $\checkmark$ \\ 
\verb|0.411_Tb5Ge4.mcif| & &  &  & $\checkmark$ & $\checkmark$ & $\checkmark$ & $\checkmark$ &  &  \\ 
\verb|0.412_Tb5Ge4.mcif| & &  &  & $\checkmark$ & $\checkmark$ & $\checkmark$ & $\checkmark$ &  &  \\ 
\verb|0.419_ErGe2O7.mcif| & &  &  & $\checkmark$ & $\checkmark$ & $\checkmark$ & $\checkmark$ & $\checkmark$ & $\checkmark$ \\ 
\verb|0.42_HoMnO3.mcif| & &  &  &  &  &  & $\checkmark$ &  &  \\ 
\verb|0.430_Yb3Pt4.mcif| & &  &  & $\checkmark$ & $\checkmark$ & $\checkmark$ & $\checkmark$ &  &  \\ 
\verb|0.431_CuB2O4.mcif| & & $\checkmark$ & $\checkmark$ & $\checkmark$ & $\checkmark$ & $\checkmark$ & $\checkmark$ & $\checkmark$ & $\checkmark$ \\ 
\verb|0.43_HoMnO3.mcif| & &  &  &  &  &  & $\checkmark$ &  &  \\ 
\verb|0.440_SrCuTe2O6.mcif| & &  &  & $\checkmark$ & $\checkmark$ & $\checkmark$ & $\checkmark$ & $\checkmark$ & $\checkmark$ \\ 
\verb|0.450_Nd5Ge4.mcif| &$\checkmark$ & $\checkmark$ & $\checkmark$ &  &  &  & $\checkmark$ & $\checkmark$ & $\checkmark$ \\ 
\verb|0.478_SmCrO3.mcif| & &  &  &  &  &  & $\checkmark$ & $\checkmark$ & $\checkmark$ \\ 
\verb|0.479_SmCrO3.mcif| &$\checkmark$ & $\checkmark$ & $\checkmark$ &  &  &  & $\checkmark$ & $\checkmark$ & $\checkmark$ \\ 
\verb|0.488_YbMnO3.mcif| & &  &  &  &  &  & $\checkmark$ &  &  \\ 
\verb|0.489_YbMnO3.mcif| & &  &  &  &  &  & $\checkmark$ &  &  \\ 
\verb|0.48_Tb2Sn2O7.mcif| &$\checkmark$ & $\checkmark$ & $\checkmark$ &  &  &  & $\checkmark$ & $\checkmark$ & $\checkmark$ \\ 
\verb|0.490_YbMnO3.mcif| &$\checkmark$ & $\checkmark$ & $\checkmark$ & $\checkmark$ & $\checkmark$ & $\checkmark$ & $\checkmark$ & $\checkmark$ & $\checkmark$ \\ 
\verb|0.49_Ho2Ru2O7.mcif| &$\checkmark$ & $\checkmark$ & $\checkmark$ &  &  &  & $\checkmark$ & $\checkmark$ & $\checkmark$ \\ 
\verb|0.51_Ho2Ru2O7.mcif| &$\checkmark$ & $\checkmark$ & $\checkmark$ &  &  &  & $\checkmark$ & $\checkmark$ & $\checkmark$ \\ 
\verb|0.530_SrCuTe2O6.mcif| & &  &  & $\checkmark$ & $\checkmark$ & $\checkmark$ & $\checkmark$ & $\checkmark$ & $\checkmark$ \\ 
\verb|0.544_Mn2FeReO6.mcif| &$\checkmark$ & $\checkmark$ & $\checkmark$ &  &  &  & $\checkmark$ & $\checkmark$ & $\checkmark$ \\ 
\verb|0.545_Mn2FeReO6.mcif| &$\checkmark$ & $\checkmark$ & $\checkmark$ &  &  &  & $\checkmark$ & $\checkmark$ & $\checkmark$ \\ 
\verb|0.571_CoSO4.mcif| & &  &  &  &  &  & $\checkmark$ & $\checkmark$ & $\checkmark$ \\ 
\verb|0.572_Na2NiCrF7.mcif| &$\checkmark$ & $\checkmark$ & $\checkmark$ &  &  &  & $\checkmark$ & $\checkmark$ & $\checkmark$ \\ 
\verb|0.573_Na2NiCrF7.mcif| &$\checkmark$ & $\checkmark$ & $\checkmark$ &  &  &  & $\checkmark$ & $\checkmark$ & $\checkmark$ \\ 
\verb|0.574_MnFeF5(H2O)2.mcif| &$\checkmark$ & $\checkmark$ & $\checkmark$ & $\checkmark$ & $\checkmark$ & $\checkmark$ & $\checkmark$ & $\checkmark$ & $\checkmark$ \\ 
\verb|0.576_Cr2F5.mcif| &$\checkmark$ & $\checkmark$ & $\checkmark$ &  &  &  & $\checkmark$ & $\checkmark$ & $\checkmark$ \\ 
\verb|0.578_NaBaFe2F9.mcif| &$\checkmark$ & $\checkmark$ & $\checkmark$ &  &  &  & $\checkmark$ & $\checkmark$ & $\checkmark$ \\ 
\verb|0.584_Fe2F5(H2O)2.mcif| &$\checkmark$ & $\checkmark$ & $\checkmark$ &  &  &  & $\checkmark$ & $\checkmark$ & $\checkmark$ \\ 
\verb|0.60_[NH2(CH3)2]n[FeIIIFeII(HCOO)6]n.mcif| &$\checkmark$ & $\checkmark$ & $\checkmark$ &  &  &  & $\checkmark$ & $\checkmark$ & $\checkmark$ \\ 
\verb|0.64_MnV2O4.mcif| &$\checkmark$ & $\checkmark$ & $\checkmark$ &  &  &  & $\checkmark$ & $\checkmark$ & $\checkmark$ \\ 
\verb|0.652_HoMnO3.mcif| & &  &  &  &  &  & $\checkmark$ &  &  \\ 
\verb|0.658_BaCuTe2O6.mcif| & &  &  &  &  &  & $\checkmark$ &  &  \\ 
\verb|0.696_SmCrO3.mcif| &$\checkmark$ & $\checkmark$ & $\checkmark$ &  &  &  & $\checkmark$ & $\checkmark$ & $\checkmark$ \\ 
\verb|0.697_SmCrO3.mcif| &$\checkmark$ & $\checkmark$ & $\checkmark$ &  &  &  & $\checkmark$ & $\checkmark$ & $\checkmark$ \\ 
\verb|0.70_Na3Co(CO3)2Cl.mcif| & &  &  &  &  &  & $\checkmark$ & $\checkmark$ & $\checkmark$ \\ 
\verb|0.715_HoCrWO6.mcif| & &  &  & $\checkmark$ & $\checkmark$ & $\checkmark$ & $\checkmark$ & $\checkmark$ & $\checkmark$ \\ 
\verb|0.726_CsMn2F6.mcif| &$\checkmark$ & $\checkmark$ & $\checkmark$ &  &  &  & $\checkmark$ & $\checkmark$ & $\checkmark$ \\ 
\verb|0.727_CsMn2F6.mcif| &$\checkmark$ & $\checkmark$ & $\checkmark$ &  &  &  & $\checkmark$ & $\checkmark$ & $\checkmark$ \\ 
\verb|0.740_Dy3Ga5O12.mcif| & &  &  &  &  &  & $\checkmark$ &  &  \\ 
\verb|0.741_Er3Ga5O12.mcif| & &  &  &  &  &  & $\checkmark$ &  &  \\ 
\verb|0.743_Ho3Al5O12.mcif| & &  &  &  &  &  & $\checkmark$ &  &  \\ 
\verb|0.744_Tb3Al5O12.mcif| & &  &  &  &  &  & $\checkmark$ &  &  \\ 
\verb|0.745_Ho3Ga5O12.mcif| & &  &  &  &  &  & $\checkmark$ &  &  \\ 
\verb|0.746_Tb3Ga5O12.mcif| & &  &  &  &  &  & $\checkmark$ &  &  \\ 
\verb|0.756_GaV4S8.mcif| &$\checkmark$ & $\checkmark$ & $\checkmark$ & $\checkmark$ & $\checkmark$ & $\checkmark$ & $\checkmark$ & $\checkmark$ & $\checkmark$ \\ 
\verb|0.763_Mn5(PO4)2(PO3(OH))2(HOH)4.mcif| &$\checkmark$ & $\checkmark$ & $\checkmark$ &  &  &  & $\checkmark$ & $\checkmark$ & $\checkmark$ \\ 
\verb|0.764_Mn5(PO4)2(PO3(OH))2(HOH)4.mcif| &$\checkmark$ & $\checkmark$ & $\checkmark$ &  &  &  & $\checkmark$ & $\checkmark$ & $\checkmark$ \\ 
\verb|0.765_Mn5(PO4)2(PO3(OH))2(HOH)4.mcif| &$\checkmark$ & $\checkmark$ & $\checkmark$ &  &  &  & $\checkmark$ & $\checkmark$ & $\checkmark$ \\ 
\verb|0.77_Tb2Ti2O7.mcif| &$\checkmark$ & $\checkmark$ & $\checkmark$ &  &  &  & $\checkmark$ & $\checkmark$ & $\checkmark$ \\ 
\verb|0.78_NiN2O6.mcif| &$\checkmark$ & $\checkmark$ & $\checkmark$ &  &  &  & $\checkmark$ & $\checkmark$ & $\checkmark$ \\ 
\verb|0.806_Fe2Se2O7.mcif| & &  &  & $\checkmark$ & $\checkmark$ & $\checkmark$ & $\checkmark$ &  &  \\ 
\verb|0.807_Fe2Se2O7.mcif| & &  &  & $\checkmark$ & $\checkmark$ & $\checkmark$ & $\checkmark$ &  &  \\ 
\verb|0.808_Fe2Se2O7.mcif| & &  &  & $\checkmark$ & $\checkmark$ & $\checkmark$ & $\checkmark$ &  &  \\ 
\verb|0.809_Fe2WO6.mcif| & &  &  & $\checkmark$ & $\checkmark$ & $\checkmark$ & $\checkmark$ &  &  \\ 
\verb|0.851_C7H14NFeCl4.mcif| &$\checkmark$ & $\checkmark$ & $\checkmark$ & $\checkmark$ & $\checkmark$ & $\checkmark$ & $\checkmark$ & $\checkmark$ & $\checkmark$ \\ 
\verb|0.862_Eu2Ir2O7.mcif| & &  &  &  &  &  & $\checkmark$ &  &  \\ 
\verb|0.870_Pr2NiIrO6.mcif| &$\checkmark$ & $\checkmark$ & $\checkmark$ &  &  &  & $\checkmark$ & $\checkmark$ & $\checkmark$ \\ 
\verb|0.874_Nd2NiIrO6.mcif| &$\checkmark$ & $\checkmark$ & $\checkmark$ &  &  &  & $\checkmark$ & $\checkmark$ & $\checkmark$ \\ 
\verb|0.875_Nd2NiIrO6.mcif| &$\checkmark$ & $\checkmark$ & $\checkmark$ &  &  &  & $\checkmark$ & $\checkmark$ & $\checkmark$ \\ 
\verb|0.877_Nd2ZnIrO6.mcif| & &  &  &  &  &  & $\checkmark$ &  &  \\ 
\verb|0.878_Nd2ZnIrO6.mcif| & &  &  &  &  &  & $\checkmark$ &  &  \\ 
\verb|0.879_Nd2ZnIrO6.mcif| & &  &  &  &  &  & $\checkmark$ &  &  \\ 
\verb|0.883_NaCo2(SeO3)2(OH).mcif| &$\checkmark$ & $\checkmark$ & $\checkmark$ &  &  &  & $\checkmark$ & $\checkmark$ & $\checkmark$ \\ 
\verb|0.898_Mn3IrSi.mcif| & &  &  & $\checkmark$ & $\checkmark$ & $\checkmark$ & $\checkmark$ & $\checkmark$ & $\checkmark$ \\ 
\verb|0.899_Mn3IrGe.mcif| & &  &  & $\checkmark$ & $\checkmark$ & $\checkmark$ & $\checkmark$ & $\checkmark$ & $\checkmark$ \\ 
\verb|0.900_Mn3CoGe.mcif| & &  &  & $\checkmark$ & $\checkmark$ & $\checkmark$ & $\checkmark$ & $\checkmark$ & $\checkmark$ \\ 
\verb|0.90_Rb2Fe2O(AsO4)2.mcif| & &  &  &  &  &  & $\checkmark$ & $\checkmark$ & $\checkmark$ \\ 
\verb|0.916_Cd2Os2O7.mcif| & &  &  &  &  &  & $\checkmark$ &  &  \\ 
\verb|0.91_Rb2Fe2O(AsO4)2.mcif| &$\checkmark$ & $\checkmark$ & $\checkmark$ &  &  &  & $\checkmark$ & $\checkmark$ & $\checkmark$ \\ 
\verb|0.941_Er2O3.mcif| & &  &  &  &  &  & $\checkmark$ & $\checkmark$ & $\checkmark$ \\ 
\verb|0.942_Er2Ge2O7.mcif| & &  &  & $\checkmark$ & $\checkmark$ & $\checkmark$ & $\checkmark$ & $\checkmark$ & $\checkmark$ \\ 
\verb|0.943_Yb2Ge2O7.mcif| &$\checkmark$ & $\checkmark$ & $\checkmark$ & $\checkmark$ & $\checkmark$ & $\checkmark$ & $\checkmark$ & $\checkmark$ & $\checkmark$ \\ 
\verb|0.944_Yb2Ir2O7.mcif| &$\checkmark$ & $\checkmark$ & $\checkmark$ &  &  &  & $\checkmark$ & $\checkmark$ & $\checkmark$ \\ 
\verb|0.945_Yb2Ir2O7.mcif| & &  &  &  &  &  & $\checkmark$ &  &  \\ 
\verb|0.948_CaNi3P4O14.mcif| &$\checkmark$ & $\checkmark$ & $\checkmark$ &  &  &  & $\checkmark$ & $\checkmark$ & $\checkmark$ \\ 
\verb|0.950_LaErO3.mcif| & &  &  &  &  &  & $\checkmark$ & $\checkmark$ & $\checkmark$ \\ 
\verb|0.954_Nd2Ir2O7.mcif| & &  &  &  &  &  & $\checkmark$ &  &  \\ 
\verb|0.958_Mn3Si2Te6.mcif| &$\checkmark$ & $\checkmark$ & $\checkmark$ &  &  &  & $\checkmark$ & $\checkmark$ & $\checkmark$ \\ 
\verb|0.96_CoSO4.mcif| & &  &  &  &  &  & $\checkmark$ & $\checkmark$ & $\checkmark$ \\ 
\verb|0.97_FeSb2O4.mcif| & &  &  &  &  & $\checkmark$ &  &  & $\checkmark$ \\ 
\verb|1.0.23_Dy3Ru4Al12.mcif| &$\checkmark$ & $\checkmark$ & $\checkmark$ &  &  &  & $\checkmark$ & $\checkmark$ & $\checkmark$ \\ 
\verb|1.0.52_Tb14Ag51.mcif| & &  &  & $\checkmark$ & $\checkmark$ & $\checkmark$ & $\checkmark$ &  &  \\ 
\verb|1.102_U2Ni2In.mcif| & &  &  &  &  &  & $\checkmark$ &  &  \\ 
\verb|1.115_Dy3Ru4Al12.mcif| & &  &  &  &  &  & $\checkmark$ &  &  \\ 
\verb|1.135_C8H10Co2O11.mcif| & &  &  &  &  &  & $\checkmark$ &  &  \\ 
\verb|1.138_MgV2O4.mcif| & &  &  &  &  &  & $\checkmark$ &  &  \\ 
\verb|1.161_PrFe3(BO3)4.mcif| & &  &  &  &  &  & $\checkmark$ &  &  \\ 
\verb|1.167_NiS2.mcif| & &  &  &  &  &  & $\checkmark$ &  &  \\ 
\verb|1.201_Cr2ReO6.mcif| & &  &  &  &  &  & $\checkmark$ &  &  \\ 
\verb|1.207_U2Rh2Sn.mcif| & &  &  &  &  &  & $\checkmark$ &  &  \\ 
\verb|1.235_Ba(TiO)Cu4(PO4)4.mcif| & &  &  &  &  &  & $\checkmark$ &  &  \\ 
\verb|1.267_Dy2Co3Al9.mcif| & &  &  &  &  &  & $\checkmark$ &  &  \\ 
\verb|1.274_DyFeWO6.mcif| & &  &  &  &  &  & $\checkmark$ &  &  \\ 
\verb|1.279_Ho2Cu2O5.mcif| & &  &  &  &  &  & $\checkmark$ &  &  \\ 
\verb|1.299_GdMn2O5.mcif| & &  &  &  &  &  & $\checkmark$ &  &  \\ 
\verb|1.300_GdMn2O5.mcif| & &  &  &  &  &  & $\checkmark$ &  &  \\ 
\verb|1.303_Dy3Ru4Al12.mcif| & &  &  &  &  &  & $\checkmark$ &  &  \\ 
\verb|1.307_Mn5Si3.mcif| & &  &  &  &  &  &  &  &  \\ 
\verb|1.326_PrMn2O5.mcif| & &  &  &  &  &  & $\checkmark$ &  &  \\ 
\verb|1.327_LaMn2O5.mcif| & &  &  &  &  &  & $\checkmark$ &  &  \\ 
\verb|1.342_Co3(PO4)2.mcif| & &  &  &  &  &  & $\checkmark$ &  &  \\ 
\verb|1.498_Cu6(SiO3)6(H2O)6.mcif| & &  &  &  &  &  & $\checkmark$ &  &  \\ 
\verb|1.595_CaCoSO.mcif| & &  &  &  &  &  & $\checkmark$ &  &  \\ 
\verb|1.680_Nd2NiIrO6.mcif| & &  &  &  &  &  & $\checkmark$ &  &  \\ 
\verb|1.710_BaFe2Se3.mcif| & &  &  &  &  &  & $\checkmark$ &  &  \\ 
\verb|1.720_Yb2O3.mcif| & &  &  &  &  &  & $\checkmark$ &  &  \\ 
\verb|1.73_CaV2O4.mcif| & &  &  &  &  &  & $\checkmark$ &  &  \\ 
\verb|1.75_BiMn2O5.mcif| & &  &  &  &  &  & $\checkmark$ &  &  \\ 
\verb|1.85_alpha-Mn.mcif| & &  &  &  &  &  & $\checkmark$ &  &  \\ 
\verb|1.89_DyFe3(BO3)4.mcif| & &  &  &  &  &  & $\checkmark$ &  &  \\ 
\verb|1.92_HoFe3(BO3)4.mcif| & &  &  &  &  &  & $\checkmark$ &  &  \\ 
\verb|2.18_Sc2NiMnO6.mcif| & &  &  &  &  &  & $\checkmark$ &  &  \\ 
\verb|2.19_Mn3ZnC.mcif| &$\checkmark$ & $\checkmark$ & $\checkmark$ &  &  &  & $\checkmark$ &  & $\checkmark$ \\ 
\verb|2.32_Dy3Ru4Al12.mcif| &$\checkmark$ & $\checkmark$ & $\checkmark$ &  &  &  & $\checkmark$ & $\checkmark$ & $\checkmark$ \\ 
\verb|2.33_Na2Mn3Se4.mcif| & &  &  &  &  &  & $\checkmark$ &  &  \\ 
\verb|2.35_CrSe.mcif| & &  & $\checkmark$ &  & $\checkmark$ & $\checkmark$ &  &  & $\checkmark$ \\ 
\verb|2.37_La8Cu7O19.mcif| & &  &  &  &  &  & $\checkmark$ &  &  \\ 
\verb|2.38_Pb2MnWO6.mcif| & &  &  & $\checkmark$ & $\checkmark$ & $\checkmark$ & $\checkmark$ & $\checkmark$ & $\checkmark$ \\ 
\verb|2.3_HoNiO3.mcif| &$\checkmark$ & $\checkmark$ & $\checkmark$ & $\checkmark$ & $\checkmark$ & $\checkmark$ & $\checkmark$ & $\checkmark$ & $\checkmark$ \\ 
\verb|2.52_Mn3O4.mcif| &$\checkmark$ & $\checkmark$ & $\checkmark$ & $\checkmark$ & $\checkmark$ & $\checkmark$ & $\checkmark$ & $\checkmark$ & $\checkmark$ \\ 
\verb|2.55_Sr2Fe3Se2O3.mcif| & &  &  &  &  &  & $\checkmark$ &  &  \\ 
\verb|2.59_Mn3As2.mcif| &$\checkmark$ & $\checkmark$ & $\checkmark$ &  &  &  & $\checkmark$ & $\checkmark$ & $\checkmark$ \\ 
\verb|2.5_Mn3CuN.mcif| &$\checkmark$ & $\checkmark$ & $\checkmark$ &  &  &  & $\checkmark$ &  & $\checkmark$ \\ 
\verb|2.61_Fe3F8(H2O)2.mcif| & & $\checkmark$ & $\checkmark$ &  &  &  & $\checkmark$ & $\checkmark$ & $\checkmark$ \\ 
\verb|2.62_TbCrO3.mcif| & & $\checkmark$ & $\checkmark$ & $\checkmark$ & $\checkmark$ & $\checkmark$ & $\checkmark$ & $\checkmark$ & $\checkmark$ \\ 
\verb|2.63_DyCrO3.mcif| & & $\checkmark$ & $\checkmark$ &  &  &  & $\checkmark$ & $\checkmark$ & $\checkmark$ \\ 
\verb|2.64_DyCrO3.mcif| & & $\checkmark$ & $\checkmark$ &  &  &  & $\checkmark$ & $\checkmark$ & $\checkmark$ \\ 
\verb|2.75_Sr2Fe3S2O3.mcif| & &  &  &  &  &  & $\checkmark$ &  &  \\ 
\verb|2.76_Sr2Fe3Se2O3.mcif| & &  &  &  &  &  & $\checkmark$ &  &  \\ 
\verb|2.91_NaCo2(SeO3)2(OH).mcif| &$\checkmark$ & $\checkmark$ & $\checkmark$ & $\checkmark$ & $\checkmark$ & $\checkmark$ & $\checkmark$ &  & $\checkmark$ \\ 
\verb|2.9_Ca3CuNi2(PO4)4.mcif| & &  &  &  &  &  & $\checkmark$ &  &  \\ 
\verb|3.10_NpSe.mcif| & &  &  &  &  &  & $\checkmark$ &  &  \\ 
\verb|3.11_NpTe.mcif| & &  &  &  &  &  & $\checkmark$ &  &  \\ 
\verb|3.12_USb.mcif| & &  &  &  &  &  &  &  &  \\ 
\verb|3.16_Gd2Ti2O7.mcif| & &  &  &  &  &  & $\checkmark$ &  &  \\ 
\verb|3.18_HoRh.mcif| & &  &  &  &  &  &  &  &  \\ 
\verb|3.19_CoO.mcif| & &  &  &  &  &  & $\checkmark$ &  &  \\ 
\verb|3.2_UO2.mcif| & &  &  &  &  &  &  &  &  \\ 
\verb|3.4_MgCr2O4.mcif| & & $\checkmark$ & $\checkmark$ &  & $\checkmark$ & $\checkmark$ &  &  & $\checkmark$ \\ 
\verb|3.6_DyCu.mcif| & &  &  &  &  &  &  &  &  \\ 
\verb|3.7_NpBi.mcif| & &  &  &  &  &  &  &  &  \\ 
\verb|3.8_NdZn.mcif| & &  &  &  &  &  &  &  &  \\ 
\verb|3.9_NpS.mcif| & &  &  &  &  &  & $\checkmark$ &  &  \\ 
\bottomrule
\end{longtable*}

\clearpage

\end{document}